\documentclass[10pt,letterpaper]{article}
\usepackage[top=0.85in,left=1in, footskip=0.75in,marginparwidth=2in]{geometry}

\usepackage[utf8]{inputenc}

\usepackage{cite}

\usepackage{nameref,hyperref}

\usepackage[right]{lineno}

\usepackage{microtype}
\DisableLigatures[f]{encoding = *, family = * }

\raggedright
\usepackage{changepage}

\usepackage[aboveskip=1pt,labelfont=bf,labelsep=period,singlelinecheck=off]{caption}

\makeatletter
\renewcommand{\@biblabel}[1]{\quad#1.}
\makeatother

\usepackage{lastpage,fancyhdr,graphicx}
\usepackage{epstopdf}
\fancyheadoffset[L]{2.25in}
\fancyfootoffset[L]{2.25in}

\usepackage{color}

\definecolor{Gray}{gray}{.25}

\usepackage{graphicx}

\usepackage{sidecap}

\usepackage{booktabs}
\usepackage{algorithm}
\usepackage{algpseudocode}
\usepackage{multirow}
\usepackage{amsmath}

\begin{document}
\vspace*{0.35in}

\begin{flushleft}
{\Large
\textbf\newline{On the use of satellite information to estimate agricultural carbon footprint in a small area framework.}
}
\newline
\\
Riccardo Pajno\textsuperscript{1,*},
Felicetta Carillo\textsuperscript{2},
Paolo Maranzano\textsuperscript{1,3},
Timo Schmid\textsuperscript{4},
Riccardo Borgoni\textsuperscript{1},
\\
\bigskip
\bf{1} \small{Department of Economics, Management and Statistics (DEMS), University of Milano-Bicocca, Piazza dell'Ateneo Nuovo 1, Milan, 20126, Italy}
\\
\bf{2} \small{Centro di ricerca Politiche e Bioeconomia, Consiglio per la ricerca in agricoltura e l'analisi dell'economia agraria (CREA-PB), Via Giacomo Venezian 26, Milan, 20133, Italy}
\\
\bf{3} \small{Fondazione Eni Enrico Mattei (FEEM), Corso Magenta 63, Milan, 20123, Italy}\\
\bf{4} \small{Institute of Statistics, University of Bamberg, Feldkirchenstraße 21, Bamberg, 96045, Germany}\\
\bigskip
* r.pajno@campus.unimib.it

\end{flushleft}

\section*{Abstract}
The agricultural sector is undergoing rapid change due to climate pressures, demographic shifts, and uneven economic development, increasing the demand for reliable environmental indicators at fine spatial scales. However, limited data availability often constrains subregional analyses. This study develops a model-based framework for producing reliable small-area estimates for assessing the agricultural carbon footprint in the Po Valley (Northern Italy), a region characterized by intensive livestock farming and high environmental pressure. 
We integrate survey, census, and satellite-derived emission data into a unified framework and produce estimates at the level of Agrarian Subregions, defined as agriculturally homogeneous municipalities by the Italian National Institute of Statistics. Satellite-based ammonia emission data are incorporated as auxiliary covariates to improve precision and spatial coherence.
A key methodological contribution is the treatment of spatial misalignment between gridded satellite data and administrative boundaries. This issue is addressed through a geostatistical upscaling procedure combined with a parametric bootstrap that propagates uncertainty from the covariate construction stage to the final small-area estimates.
The results show that satellite-derived information substantially improves the accuracy and stability of carbon footprint estimates while reducing reliance on large, heterogeneous auxiliary datasets, illustrating the potential of Earth observation data in model-based environmental statistics.\\[1em]
\textbf{Keywords:} small area estimation; satellite data; spatial misalignment; upscaling; agricultural environmental impact.


\section{Introduction} 
\label{sec:intro}

The agricultural sector is undergoing a period of rapid transformation, driven by the interrelated and pervasive impacts of climate change, demographic shifts, and disparities in economic growth between territories and populations worldwide. It is incumbent upon governments and local authorities to address these challenges in order to safeguard vulnerable socio-economic systems and territories and to facilitate positive change in rural economic development. This necessitates the undertaking of small-scale studies and an analysis of the economics of the agricultural sector and the businesses operating within it. In particular, the actions of policymakers should be guided by the implementation of territorial policies that must necessarily take into account local conditions and specificities, rather than being limited to considerations at the supply chain level \cite{ali2025identifying}. 

The availability of accurate, up-to-date, and relevant data is of critical importance for this understanding, which has often constituted the primary obstacle to rigorous research and, ultimately, to effective policymaking in the agricultural sector. The scarcity of high-quality data required to examine the evolution of agricultural systems at the integrated and micro-area levels necessitates the implementation of methodologies with technical solutions that yield more precise estimates from available data and more effectively utilize complementarities between traditional and alternative data sources.

Among the newest data sources, an outstanding role is played by remote-sensing data collected every day by the satellites orbiting the Earth, as highlighted by the highly increasing use in the recent academic literature \cite{zhao2022overview}. The advent of these source represents a cornerstone also in contemporary statistics, since the development of appropriate statistical tools to handle these peculiar data is required \cite{belmonte2025approach}.

Satellite data present both advantages and limitations that should be carefully considered. In particular, their frequent updates enable both near real-time applications and historical reconstructions. For example, weather and air quality data provided by the European Centre for Medium-Range Weather Forecasts (ECMWF) are updated weekly or monthly and include records dating back to 1950\footnote{Consider, for example, weather data from the ERA-5 data program available at \href{https://cds.climate.copernicus.eu/datasets/reanalysis-era5-land?tab=overview}{https://cds.climate.copernicus.eu/datasets/reanalysis-era5-land?tab=overview}, accessed on January 28th, 2026.}. Second, the spatial coverage is usually large (i.e., it overcomes national or administrative borders), and the spatial resolution is granular \cite{PermatasariEtAl2025}. Conversely, socio-economic, industrial, energetic or environmental data collected through institutional census (e.g., population and industry census from national statistics offices or emissions inventory from environmental protection agency, \cite{MarongiuEtAl2022}) are slowly updated and refer to typical administrative subdivisions, such as municipalities or regions. Also, consider that very specific data, like population and socio-economic data for developing countries, are often not available or only produced sporadically over time \cite{NewhouseEtAl}. Nevertheless, they provide reliable values not affected by statistical design errors.

A drawback of satellite information, closely related to spatial resolution, is that data are often provided to users on regular grids (e.g., $0.1^\circ \times 0.1^\circ$ or $0.25^\circ \times 0.25^\circ$), which do not correspond to the administrative subdivisions typically used by stakeholders. This mismatch between spatial supports, known as the Change of Support Problem (COSP), has a long history in the spatial statistics literature \cite{gotway_young_2002}. The increasing availability of satellite data has, however, renewed interest in this topic in recent years. In this framework, \cite{rash2024spatial} proposed a methodology involving geographically reweighted regression coupled with inverse distance weighting to estimate the spatial distribution of soil properties in Iraq; \cite{shawky2025spatial} developed Bayesian hierarchical coregionalization model with spatially varying coefficients fitted using integrated nested Laplace approximation (INLA) and stochastic partial differential equations \cite{CamelettiEtAl2019,VillejoEtAl2023}, which outperformed traditional physical climate models in the spatio-temporal interpolation of the maximum temperature in the Nile Basin; \cite{wang2024integration} highlighted the benefit of machine learning approaches to handle satellite information in precision agriculture. Furthermore, the mismatch of spatial supports not only requires appropriate upscaling (or downscaling) procedures, but also raises important issues regarding the propagation of uncertainty in the modeling step used to reconcile the misaligned data.

The principal aim of this paper is to provide accurate and reliable estimates of the environmental impact of agricultural activities at a high level of spatial resolution, while simultaneously facilitating the development of evidence-based agricultural policies. In particular, our objective is to examine the spatial dynamics of the carbon footprint (CF) associated with agricultural farm holdings in the Po Valley (Northern Italy), which is one of the most polluted areas in Europe and where intensive livestock farms play a significant role in driving high levels of ammonia \cite{Agrimonia2023,HassounaEtAl2023,WangEtAl2021} and secondary atmospheric particulate matter \cite{WyerEtAl2022,RodeschiniEtAl2024,MarongiuEtAl2024}. In the Italian context, the most appropriate observational units are the Agrarian Subregions (ASRs), a territorial subdivision defined by the Italian National Statistics Office (ISTAT) that aggregates municipalities into internally homogeneous areas with similar agricultural and climatic characteristics\footnote{The translated definition from the agricultural statistics glossary reads "Agrarian region: consisting of groups of municipalities according to rules of territorial continuity that are homogeneous in relation to certain natural and agrarian characteristics and, subsequently, aggregated by altitude zone." \cite{istat:statagr}}. Notable examples on the international scene are Oltrepo Pavese, renowned for its viticulture, and Lomellina, celebrated for its rice production. The entire area under consideration, that is, the Po Valley, is officially divided into 254 distinct areas.

This study examines the relationship between agricultural CF, economic and techno-productive characteristics of farms (e.g., number and type of farm animals or available agricultural areas), landscape features  (e.g., orography) and emission inventories of agricultural-related atmospheric compounds for the year 2020. 

By combining survey and census data, we conduct a comparison between direct estimates of the CF and model-based estimates obtained employing small area spatial models belonging to the \cite{FH79} family, hereafter denoted as FH model. Small area estimation (SAE) models for Italian ASRs have previously been presented in other empirical studies, including those by \cite{McConvilleEtAl2023} on the relationship between manure and pollution in Lombardy and \cite{CarilloEtAl2024} on the estimation of key economic dimensions in subregional agricultural systems.

In the recent dissertation by \cite{tzavidis2025small} on SAE, the author identifies in the availability of new and alternative data sources, such as satellite observations, a unique opportunity to substantially improve the empirical analysis and a motivation for new methodological research on the topic, at the same time. Recent studies have applied SAE models incorporating such data in various domains, including the socio-economic field \cite{NewhouseEtAl, PermatasariEtAl2025, GardiniEtAl2025}, agricultural production \cite{RawatEtAl2025}, and forest monitoring \cite{GeorgakisEtAl2023, GeorgakisEtAl2024}. Nevertheless, the use of satellite data in SAE modeling remains relatively limited, particularly in the environmental context, where such data have the potential to provide valuable insights. In addition, \cite{tzavidis2025small} emphasizes that current works lack of a rigorous methodological way to estimate the additional uncertainty in the final estimates due to the data integration procedure (although prediction uncertainty is widely known to be a crucial aspect in SAE).

In this framework, our work proposes a geostatistical technique to handle the spatial misalignment between census and satellite data, combined with a tailored parametric bootstrap algorithm that accounts for the extra variability induced by the upscaling procedure in the SAE model. The proposed methodology makes it possible to compute reliable standard errors and confidence intervals for the estimated regression coefficients and to perform statistical testing when upscaled spatial covariates are incorporated into the regression models.

The ultimate objective of this study is to exploit satellite-derived regional ammonia (NH$_3$) emission data to enhance the robustness and reliability of agricultural CF predictions. Furthermore, the proposed approach illustrates how the integration of satellite information can mitigate the reliance on extensive auxiliary datasets, which are typically country-specific, heterogeneous in terms of definition and data collection design, and often subject to infrequent updates.

The remainder of the paper is structured as follows. Section \ref{sec:data} introduces the several data sources used in the empirical application and defines the area of interest for the agricultural CF estimation. Section \ref{sec:methods} describes the statistical methodologies used to obtain estimates of the CF, namely the direct estimate and the model-based estimate involving the Fay-Herriot SAE models. The aggregation procedure used to upscale regular gridded data to a custom support composed of irregular polygons (i.e., the ASRs) and the bootstrap algorithm are described in Section \ref{sec:upscale} and in Section \ref{sec:DoubleBoot} respectively. Section \ref{sec:results} presents the empirical results for the CF obtained using the direct estimator and the spatial Fay-Herriot models in the case of census auxiliary covariates and of satellite-derived auxiliary covariates. Section \ref{sec:conclusions} sums up the main contribution of the paper and provides the final remarks.

\section{Data and area of interest} \label{sec:data}
The data used in this paper come from several institutional sources, namely the Farms Accountancy Data Network (FADN) survey, the 2020 Italian National Census of Agriculture (INCA), the National Veterinary Registry Office Database (BDN Banca Dati Nazionale
dell’anagrafe Veterinario), and the Copernicus Atmosphere Monitoring Service (CAMS).
The FADN supplies comprehensive sample data (i.e., at the farm level) concerning the Italian agricultural sector, with a particular focus on the economic and productive structure, as well as the environmental impact of farms, via a representative sample of the national system. Consequently, the FADN information will serve as the foundation for the survey statistics employed in SAE models. Other sources furnish census data that will be used as auxiliary information in the same models.

We also consider some cartographic and orographic information related to the area of interest for this study and incorporated it into our analysis.
A full description of all the variables used in this paper is shown in Table \ref{tab:summarystats}.

\begin{table}
 \caption{Descriptive statistics of the key variables used in the paper.} 
 \label{tab:summarystats}
 \small
 \begin{tabular}{@{}lcccccccc@{}}
  \toprule
  & Source & Min & 1st Q & Median & Mean & 3rd Q & Max & Std. Dev. \\ 
  \midrule
  Utilized agricultural area & INCA & 96 & 4,878 & 10,981 & 14,293 & 20,775 & 67,799 & 12,329 \\
  Standard working days & INCA & 4,052 & 79,965 & 192,496 & 262,443 & 355,587 & 1,595,222 & 250,770 \\
  Total bovine heads & BDN & 34 & 1,849 & 4,552 & 14,343 & 17,080 & 202,045 & 24,115 \\
  Total swine heads & BDN &  0 & 124 & 3,410 & 29,189 & 22,091 & 585,859 & 69,724 \\ 
  Average altitude & ISTAT & -2 & 73 & 270 & 460 & 749 & 2,242 & 492 \\ 
  $NH_3$ emissions & CAMS & 0.003 & 0.337 & 0.937 & 1.308 & 1.600 & 9.758 & 1.418 \\ 
  CF (Kg CO$_2$ eq.) & FADN & -482,263 & 4,767 & 13,887 & 183,081 & 76,949 & 13,165,042 & 572,084 \\ 
  \bottomrule
 \end{tabular}
\end{table}

\subsection{Area of interest: the Italian Po Valley}
\begin{figure}
\centering
\includegraphics[width=0.4\linewidth]{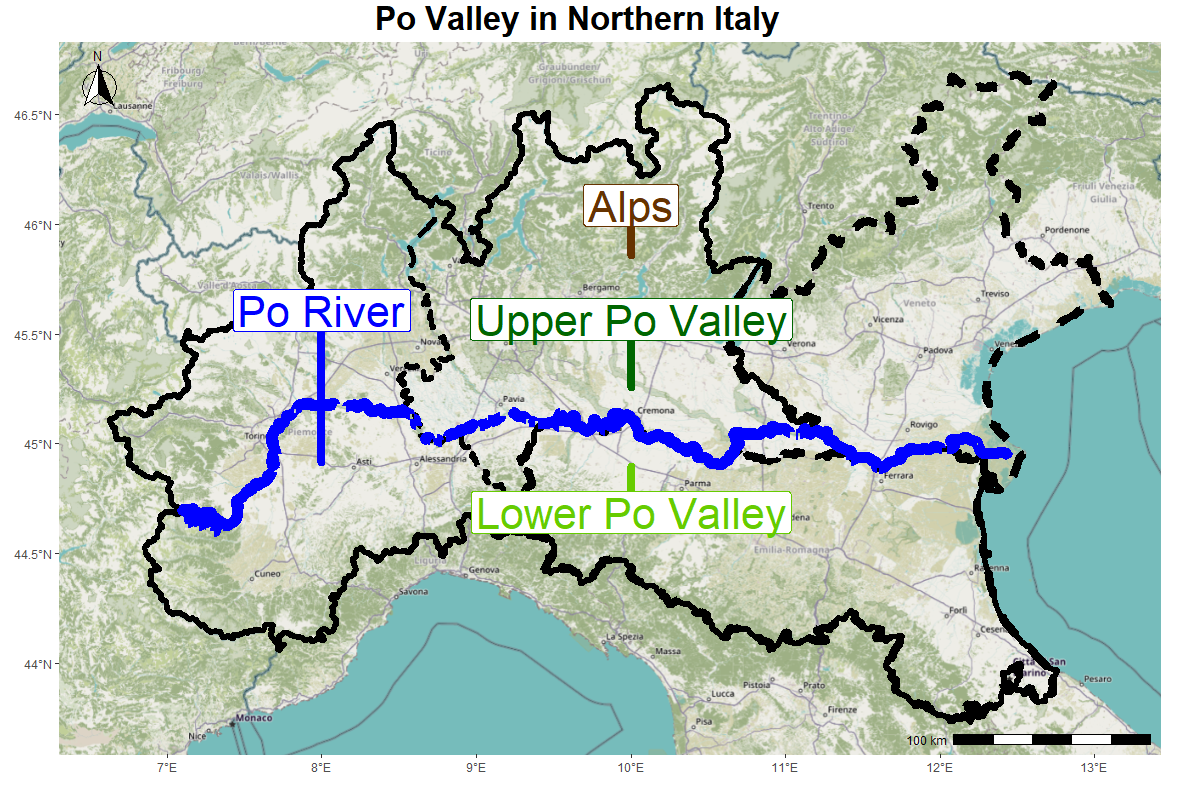}
\includegraphics[width=0.45\linewidth]{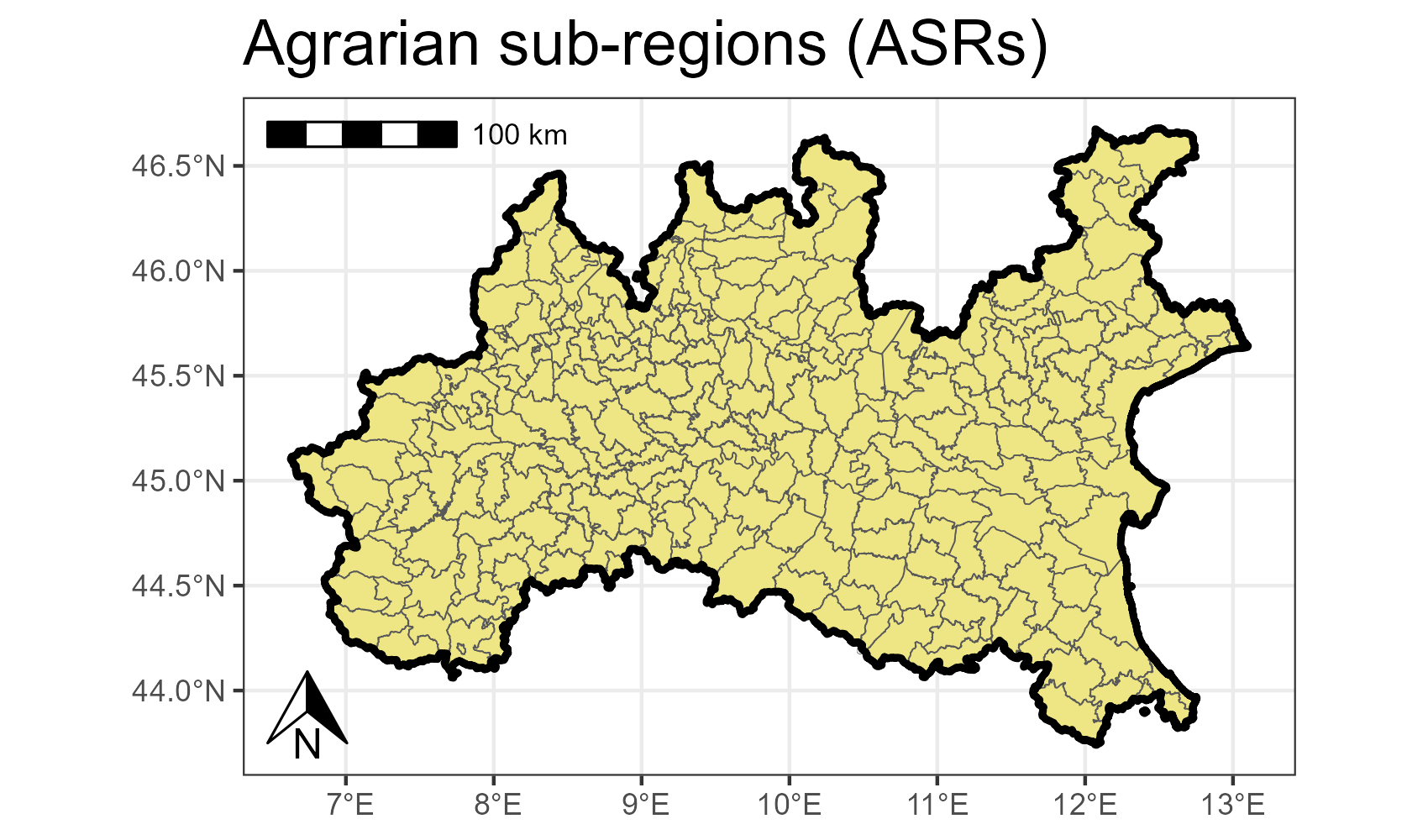}
\caption{The Po Valley and agrarian sub-regions}
\label{fig:PoValley}
\end{figure}

The area of interest for this study is referred to as the Po Valley in Northern Italy. As shown in the left panel of Figure \ref{fig:PoValley}, the term Po Valley denotes a large area located in Northern Italy traversed by the Po River, the largest river in the country. The valley is geographically defined by the union of four Italian administrative regions, namely, Lombardy, Piedmont, Veneto, and Emilia Romagna. To reference countries’ regions for statistical and policy making purposes, the European Union (EU) has developed a classification known as Nomenclature of Territorial Units for Statistics (NUTS), that divides each EU country into several geographically nested levels \cite{NUTS}. In the paper, we consider the second level of the classification, namely the NUTS-2 level or regional level, which includes the above mentioned administrative entities. The borders of the four regions are shown on the left-hand side of the Figure \ref{fig:PoValley}.

As stated above, the final goal of this study is to provide reliable estimates of the carbon footprint (CF) for small geographical areas called Agrarian Sub-Regions (ASRs). ASRs are groups of municipalities belonging to the same province and created according to territorial contiguity rules, that can be considered homogeneous according to natural and agrarian characteristics. The total number of ASRs in the study area is 254, covering 3,572 municipalities out of approximately 7,900 in the entire country, with a total population of 23,549,080 inhabitants (according to the 2020 permanent Italian Census), about $40\%$ of the Italian population.

The surface area of the ASRs ranges from a minimum of 6.85km\(^2\) in \textit{Montagna del Verbano Occidentale} to a maximum of 982.51km\(^2\) in \textit{Alto Taro}, with an average surface area of 352.75km\(^2\). The right panel of Figure \ref{fig:PoValley} shows the map of the Po Valley decomposed into the 254 ASRs. 

The Po Valley is known to be the economic engine of the country, with many facilities and industries located throughout the territory \cite{RegStatYearbook}, as well as with many urbanized centers and heterogeneous development \cite{KercukuEtAl2023}. Despite its relatively negligible weight in relation to other economic sectors, the agricultural economy exerts a substantial environmental impact. According to Eurostat \cite{ARDECO}, in 2022, the gross value added generated by agriculture and extraction constituted between 1\% and 2\% of the total gross value added in the four regions, while employment in the sector accounted for between 1\% and 3\%. 
Nevertheless, the agricultural sector was responsible for a significant share of greenhouse gas emissions: 15.18\% in Emilia-Romagna, 8.5\% in Veneto, 17.70\% in Lombardy, and 12.21\% in Piedmont.
In particular, intensive crops and breeding are spread throughout the lowlands, providing food both for domestic consumption and export. This intensive farming activity \cite{MarongiuEtAl2022,ColomboEtAl2023,MarongiuEtAl2024,Agrimonia2023}, together with its geographical position between two mountain chains \cite{RaffaelliEtAl2020} and its high level of industrialization and urbanization \cite{RegStatYearbook} makes the Po Valley one of the most polluted areas in Europe. Given the remarkable social relevance and scale of the problem, the Po Valley has been subject to numerous studies \cite{OttoEtAl2024,RodeschiniEtAl2024} and projects \cite{MarongiuEtAl2022,Agrimonia2023} directed at identifying policy interventions that can mitigate the environmental impact of the farming and livestock sector supported by data-driven methodologies.

\subsection{The FADN dataset}
Farm-level data on CF and techno-economic specialization are provided by the European Farm Accountancy Data Network (FADN) survey, which collects detailed structural and accountancy data on EU farms and is the only harmonized European statistical source on the economic management of farms. At the Italian level, the surveys are conducted by the Italian Council for Research in Agricultural Economics (CREA) through the RICA (\textit{Rete di Informazione Contabile Agricola}, Information Network for Agricultural Accounting) database. This database has been created in accordance with Law CEE 79/56 of the EU in order to collect data about European agriculture. FADN collects unit-level information on various aspects of the farm, including structural, geographical, productive, environmental, economic, and financial factors \cite{datirica}. Within the Italian scientific panorama, the FADN survey constitutes the main source of reliable information on the agricultural sector and it is used as primary data provider for both economic \cite{CoderoniEtAl2018,CoderoniPagliacci2023,CortignaniCoderoni2022} and environmental studies \cite{BaldoniEtAl2017,BaldoniEtAl2018,CoderoniVanino2022}.

The data include information from a random sample of agrarian businesses selected from the entire population of Italian farms with a Standard Output\footnote{Standard Output measures the theoretic gross product evaluated the product of theoretic yields and theoretic prices} of no less than $8,000$ euros per year. The sample encompasses $37$\% of the Italian businesses and $90$\% of the overall Italian Standard Output. The sample is stratified (thus, it is representative) according to the following three criteria: geographical location (regions), techno-economic specialization (i.e., type of farming and cropping), and economic size. The techno-economic specialization refers to the business's productive orientation and it is articulated into three levels, that is, general, main, and specific; while economic size refers to the economic dimension according to the Standard Output. 

The FADN sample used in this study comprises 2,791 farms distributed across the four regions as follows: 825 in Emilia-Romagna, 658 in Lombardy, 641 in Piedmont, and 667 in Veneto. The number of farms per ASR ranges from 0 to 89.

Farms are stratified into three types of activity: crop, livestock, and mixed. The majority are crop farms (1,929), followed by livestock farms (754), while only a small number are classified as mixed farms (108). The prevalence of crop businesses is particularly significant in Emilia-Romagna and Lombardy, while breeding farms are more common in Piedmont and Veneto.

Regarding the size, farms are classified based on whether their standard output was below $50,000$ euros (small farms) or above $50,000$ euros (large farms). We observed nearly equal numbers of small and large farms (1,517 versus 1,274). Small farms are more prevalent in Emilia-Romagna, while larger farms are concentrated in Veneto.

\begin{figure}
\centering
\includegraphics[width=0.45\linewidth]{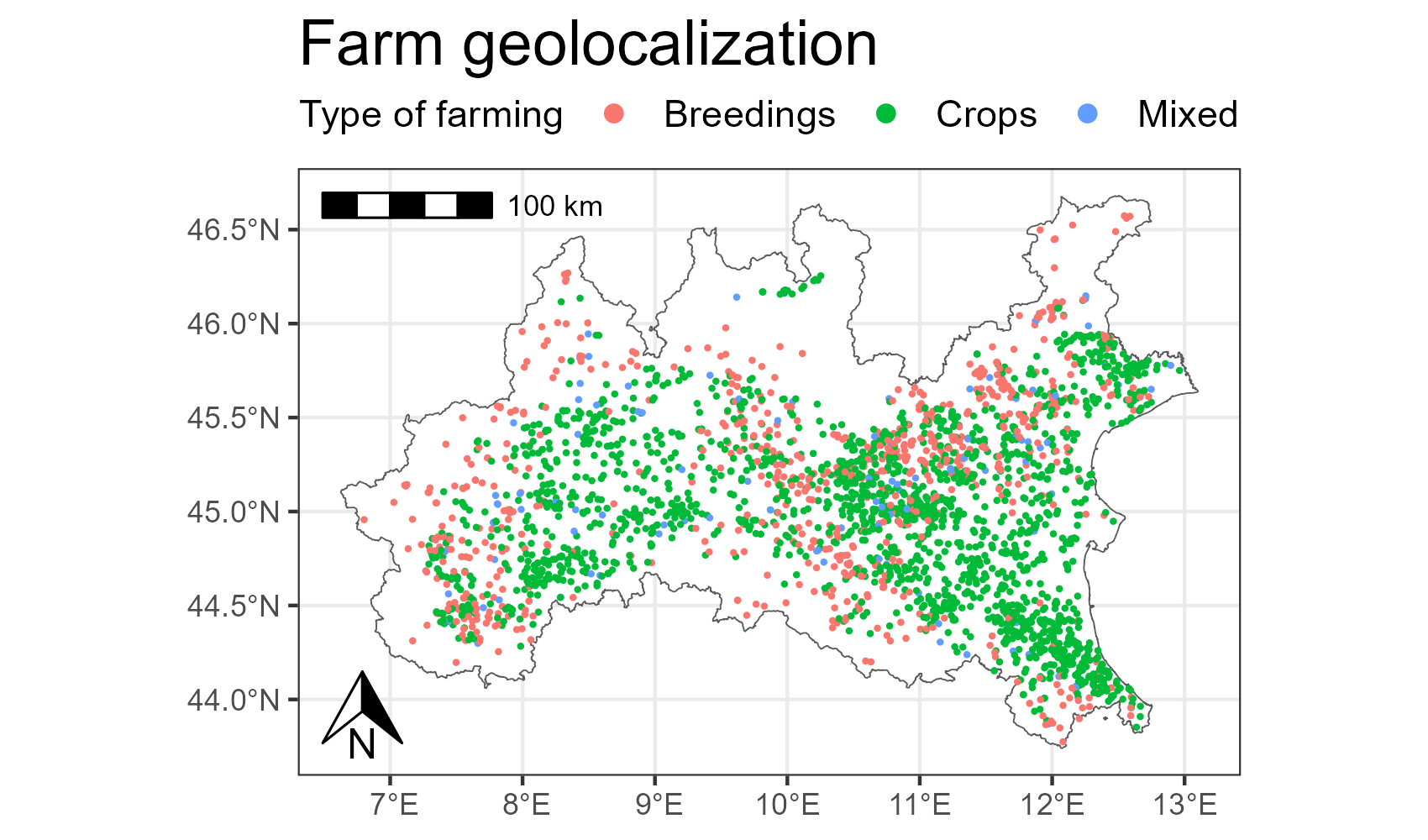}
\includegraphics[width=0.45\linewidth]{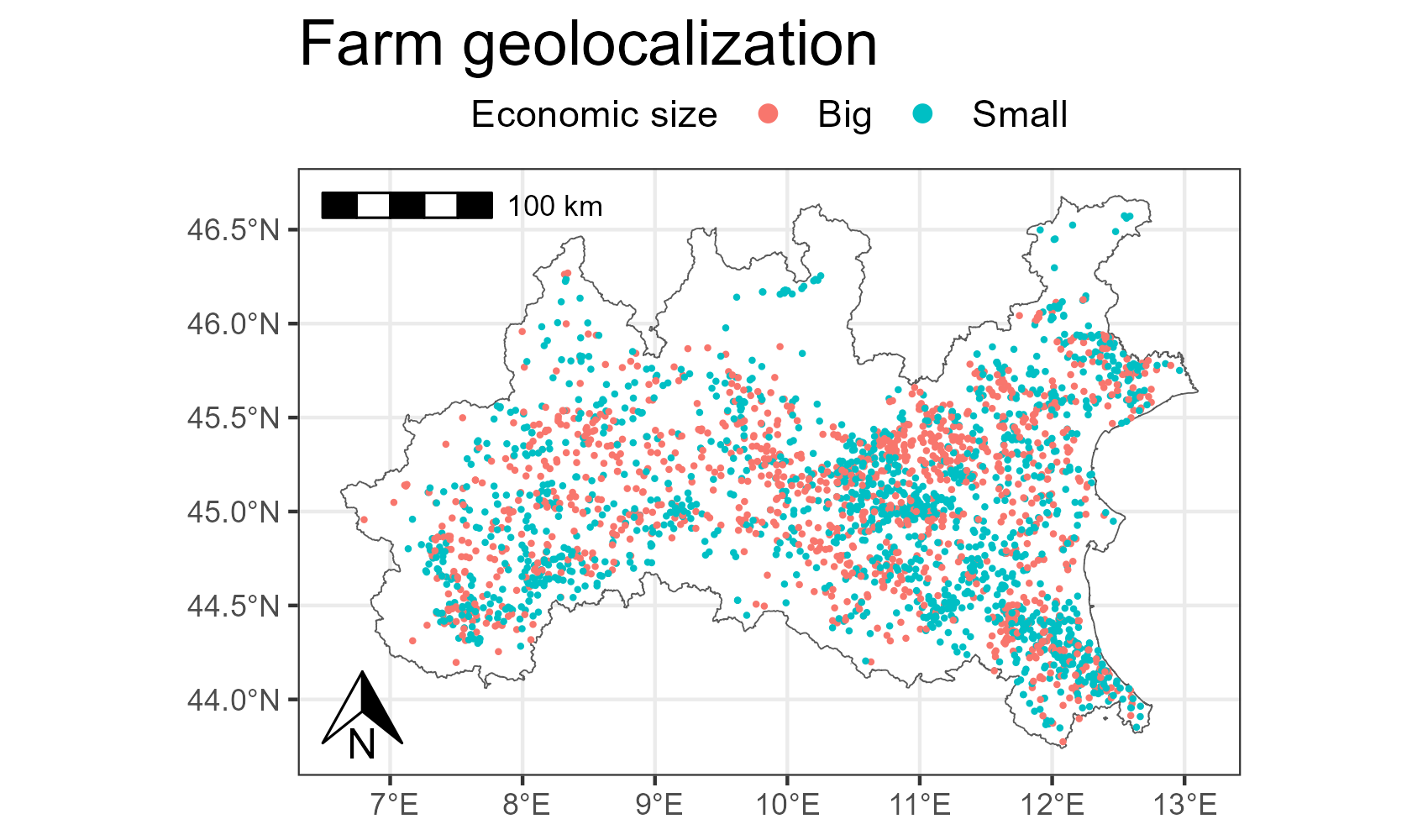}
\caption{Farm geolocalization by type of framing and economic size.}
\label{fig:farmgeo}
\end{figure}

Figure \ref{fig:farmgeo} illustrates the spatial distribution of the farms in the sample, revealing a non-uniform distribution throughout the region, with most farms concentrated in the fertile flat areas of the central Po Valley. In particular, the northern and southern mountainous regions mostly feature small-scale livestock operations. Additionally, the eastern part of the Po Valley shows a clear predominance of crop farming over livestock, whereas the western part exhibits a more balanced distribution between the two types of agriculture.

\begin{figure}
\centering
\includegraphics[width=0.45\linewidth]{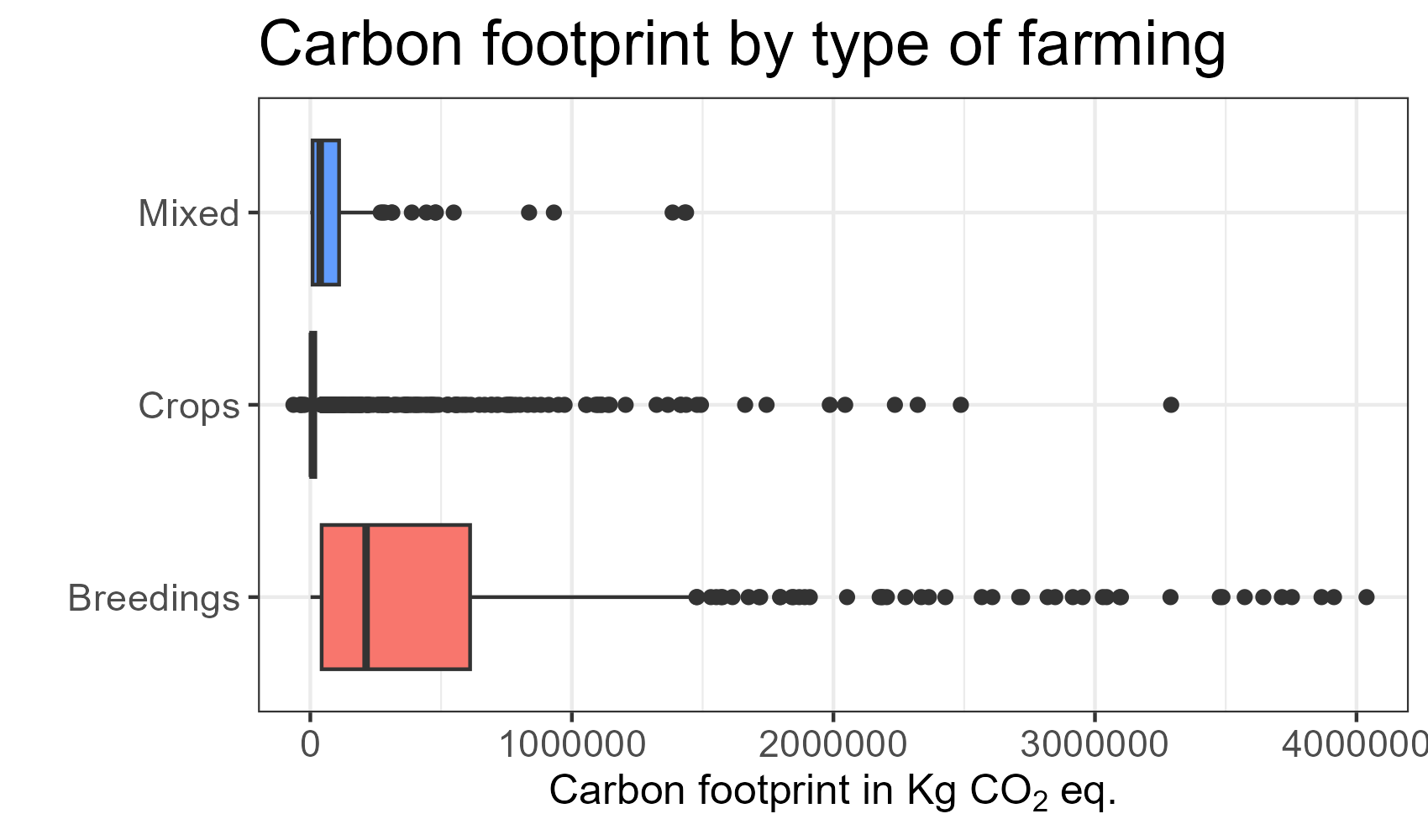}
\includegraphics[width=0.45\linewidth]{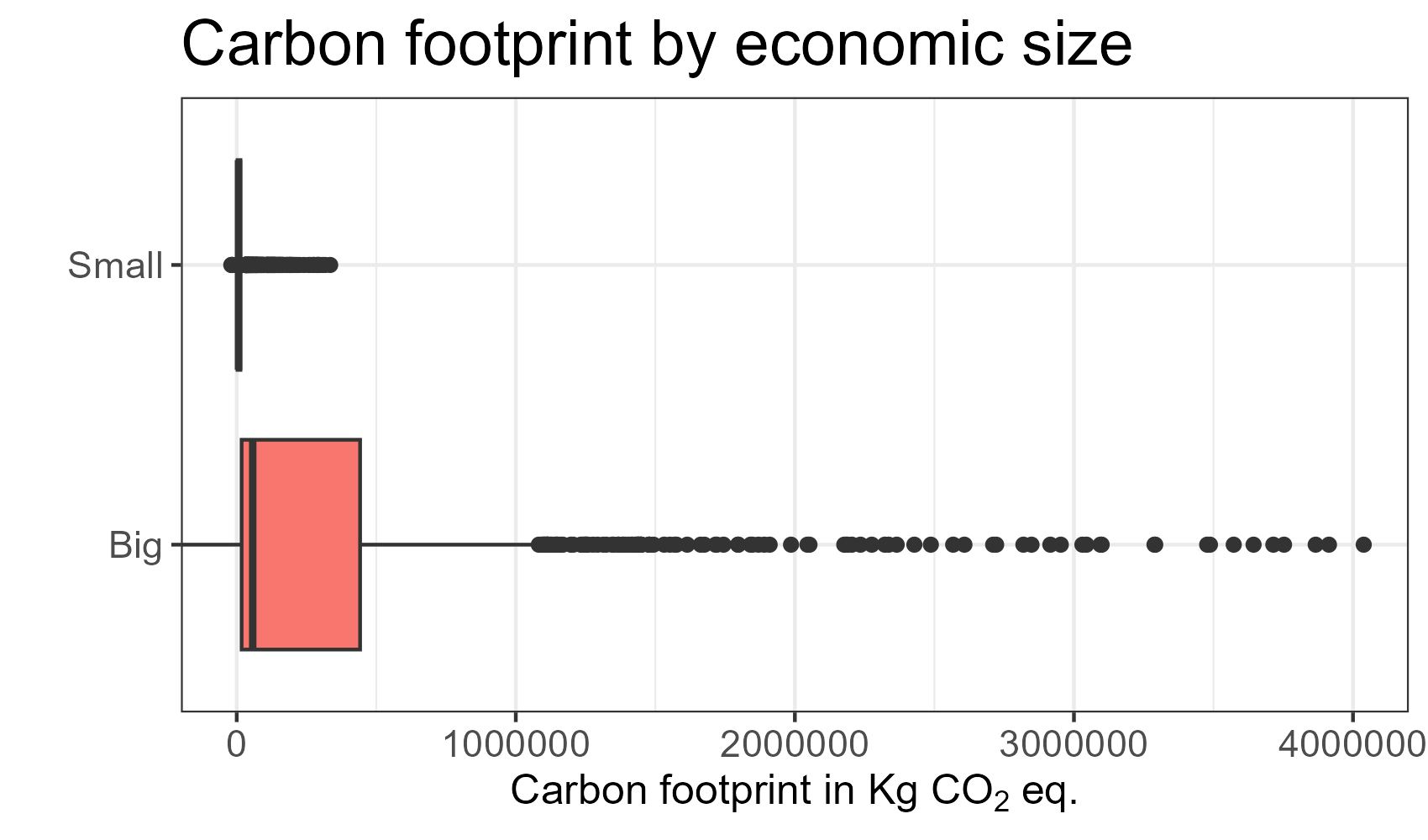}
\caption{Carbon footprint distribution conditioned to type of farming and economic size.}
\label{fig:CFbySIZE}
\end{figure}

A farm specific emission indicator of greenhouse gases (GHGs) intensity, namely the carbon footprint (CF) indicator, is included in the Italian FADN dataset. This variable has been estimated starting from micro data on the production techniques adopted on the farm and using conversion rates of emissions and of absorption of GHGs, taken from specific literature. In detail, to reconstruct a GHG farm balance, it was adopted the Intergovernmental Panel on Climate Change (IPCC) methodology at the farm level, using activity data connected to the main agricultural activities. Methane (CH$_4$) and nitrous oxide (N$_2$O) emissions are estimated from the following source categories: enteric fermentation, manure management, rice paddies, agricultural soil use, and stubble burning. These different farm-level GHG emissions are then summarized into a unique indicator using each GHG’s Global Warming Potential (GWP) \cite{CoderoniEtAl2013}. The conversion factors updated over time by the IPCC are used. GHG emissions are expressed in CO$_2$ equivalent and quantify the CF at farm level \cite{BaldoniEtAl2016}. Figure \ref{fig:CFbySIZE} displays the CF distribution by type of farming and economic size. As expected, livestock farming is significantly more polluting than crop farming, and larger farms have a higher CF than smaller ones. Also, the boxplot reveals a wide range and a strong positive skewness, with several large values corresponding to large, intensive breeding farms.

\subsection{The Italian National Census of Agriculture}
The Italian National Census of Agriculture (INCA) is led by the Italian National Statistics Office (ISTAT) and includes the entire population of agrarian companies on the Italian territory. It occurs every 10 years, and the last sweep, which is considered in this paper, took place in 2020. 
 
In 2020, there were 235,366 farms in the Po valley (46,893 farms in Lombardy, 51,703 in Piedmont, 83,017 in Veneto, and 53,753 in Emilia Romagna), showing a significant decrease compared to the 314,331 farms recorded in the previous census in 2010. This change reflects a percentage decrease of approximately $24.4\%$.
This decreasing trend is consistent across all four regions. In 2010, there were 54,333 farms in Lombardy, 67,148 in Piedmont, 119,384 in Veneto, and 73,466 in Emilia Romagna. 
Note that there were no significant changes in the Utilized Agricultural Area (UAA) over the decade, as the UAA decreased slightly from 3,873,000 hectares in 2010 to 3,829,000 hectares in 2020 \cite{istat:tavoleagr}.
As a result, we have witnessed an expansion in the size of businesses in the Po Valley, driven by capital inflows and the entry of investment funds. Although family-managed farms still account for 90$\%$ of the total, there has been a growth in the number of corporate farms, which are typically larger in size. For example, in Lombardy, the average UAA of a corporate farm is 40 hectares, compared to 16 hectares for a family-managed farm \cite{istat:tavoleagr}.

\subsection{The National Veterinary Registry Office Database}
The National Veterinary Registry Office Database (BDN \textit{Banca Dati Nazionale dell'anagrafe Veterinario}) offers a detailed amount of information about all the farm animals living in the entire Italian territory. This database is maintained by the Italian Ministry of Health for sanitary purposes. Animals are categorized into four groups: bovines, swine, ovine, and equines. In this study, we only account for bovines and swines due to their large predominance with respect to the other animals, thus having a remarkable and direct impact on the environment through GHG emissions. In the area considered, the total number of bovines is roughly 3.70 millions and the total number of swine is 7.48 millions, with significant differences between various ASRs. To appreciate the role of livestock in the Po Valley, consider that in 2024, the Lombardy region alone accounted for the 26.38\% and for the 47.65\% of the national population of bovines and swines, respectively, and that cumulating the four regions the shares reach 61.38\% and 84.58\% \cite{EurostatLivestock}.

\subsection{Orography}
\begin{figure}
\centering
\includegraphics[width=0.45\linewidth]{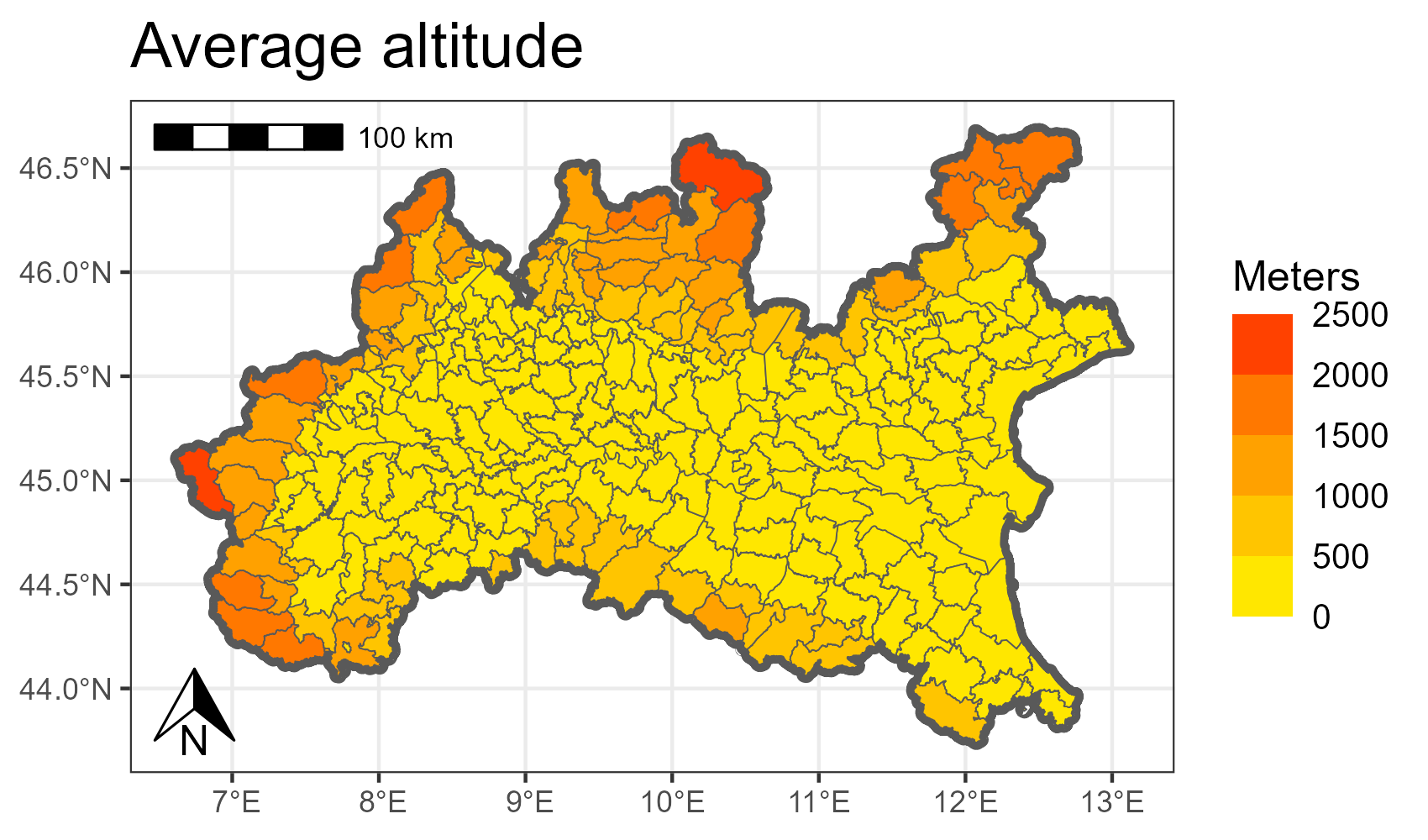}
\includegraphics[width=0.45\linewidth]{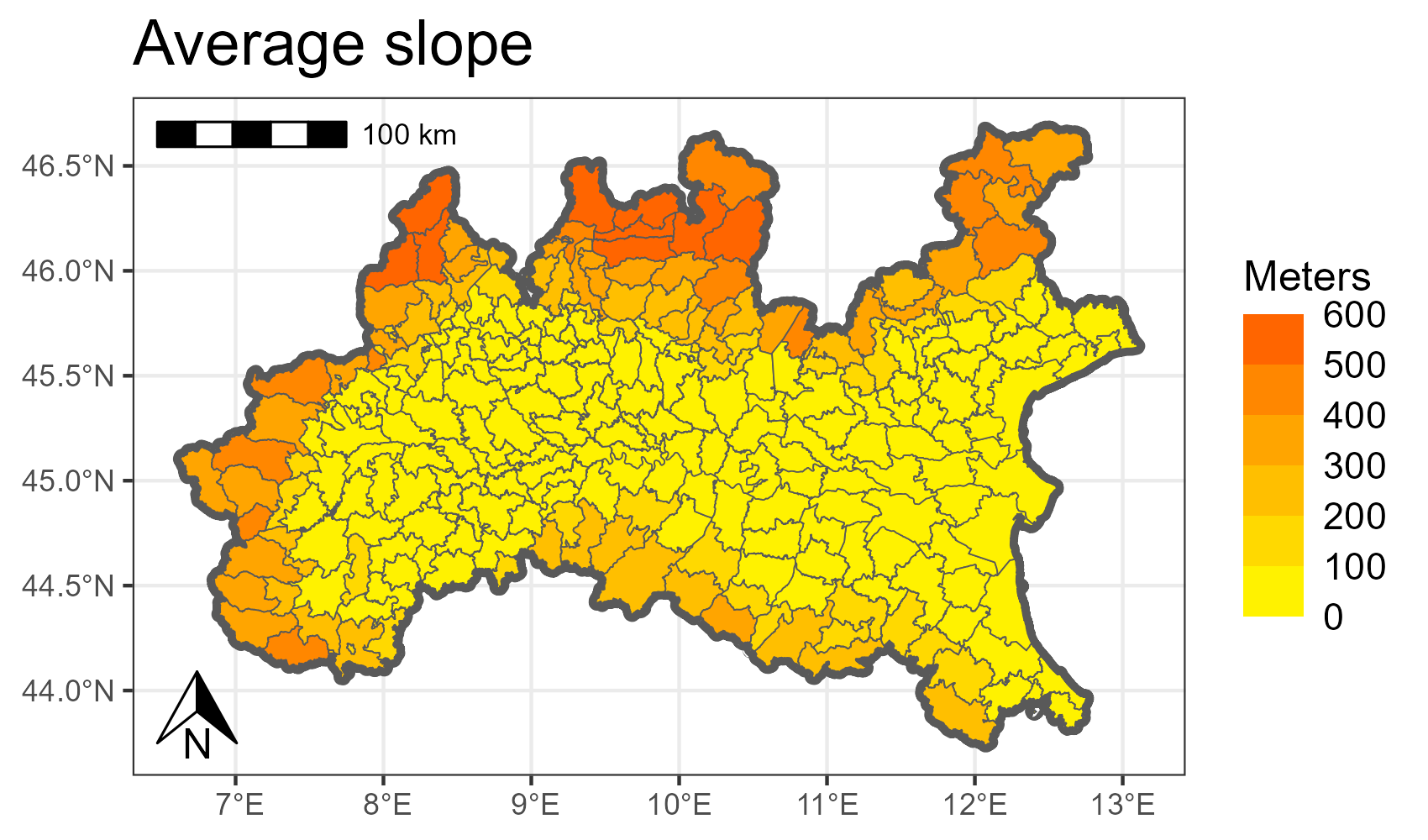}
\caption{Altitude and soil slope spatial distribution}
\label{fig:AltitudeSlope}
\end{figure}

As mentioned earlier, the Po Valley has peculiar orographic characteristics. Therefore, we additionally collected from ISTAT information about the average altitude and soil slope in order to better explain the feature of the areas and their farms. Specifically, the standard deviation of the altitude is considered a reasonable measure of the slope of the soil. Figure \ref{fig:AltitudeSlope} provides the maps of average altitude (left panel) and soil slope across the ASRs. Despite the different scales, the two maps show obvious overlapping in the values, suggesting that flat areas exhibit very few variability and weak changes in terms of altitude, while mountainous areas are characterized by large fluctuations given by the alternation of bottom valleys and the surrounding high mountains. Analytically, such a straight interrelationship is synthesized by a high degree of linear correlation, which is close to 0.93. Therefore, in the remainder of the paper, we only considered the average altitude as a proxy of the physical landscape.

\subsection{Satellite data from CAMS database}
Recent developments in spacecraft technologies created the opportunity to directly collect data on many environmental parameters, including air quality, weather, and energy use, from satellite monitoring systems. It is worth noting that, relatively to census data on the economy and the society (which are typically updated once per year or even on a larger temporal span), an enormous amount of constantly updated environmental data is being collected through remote sensing technologies. Consistently, satellite-based analysis of agriculture and land use offers large opportunities to explore new methods to efficiently make use of this information.

For Europe, the Copernicus Atmosphere Monitoring Service (CAMS) \cite{CAMS}, implemented by the European Center for Medium-Range Weather Forecasting (ECMWF) \cite{ERA5-Land}, is one of the most recent global databases of emissions and concentrations from natural and anthropogenic sources. The CAMS datasets provide monthly emission inventories for many atmospheric compounds from 2000 to 2020. These inventories are based on a combination of existing datasets and newly updated information, describing anthropogenic emissions from fossil fuel use on land, natural emissions from vegetation, soil and more. Anthropogenic land-based emissions are further disaggregated by sector of activity (e.g., transportation, agriculture).

In our case, pollutant emissions data are provided by the CAMS anthropogenic emissions dataset, which contains monthly global anthropogenic and natural emissions from 36 sources, again divided into 20 sectors (including agriculture and livestock) on a regular grid with a spatial resolution of $0.1^\circ \times 0.1^\circ$. Among others, we considered the total ammonia (NH$_3$) emissions obtained by aggregating the monthly emission values recorded in 2020. Indeed, as well established in the literature \cite{Agrimonia2023,OttoEtAl2024,RodeschiniEtAl2024}, the vast majority of the ammonia emissions are directly generated by agricultural activities, especially from livestock and manure management (as depicted in \cite{MarongiuEtAl2022,MarongiuEtAl2024} in Lombardy, agriculture contributes for roughly 97\% of the total ammonia emissions).

\begin{figure}
\centering
\includegraphics[width=10cm]{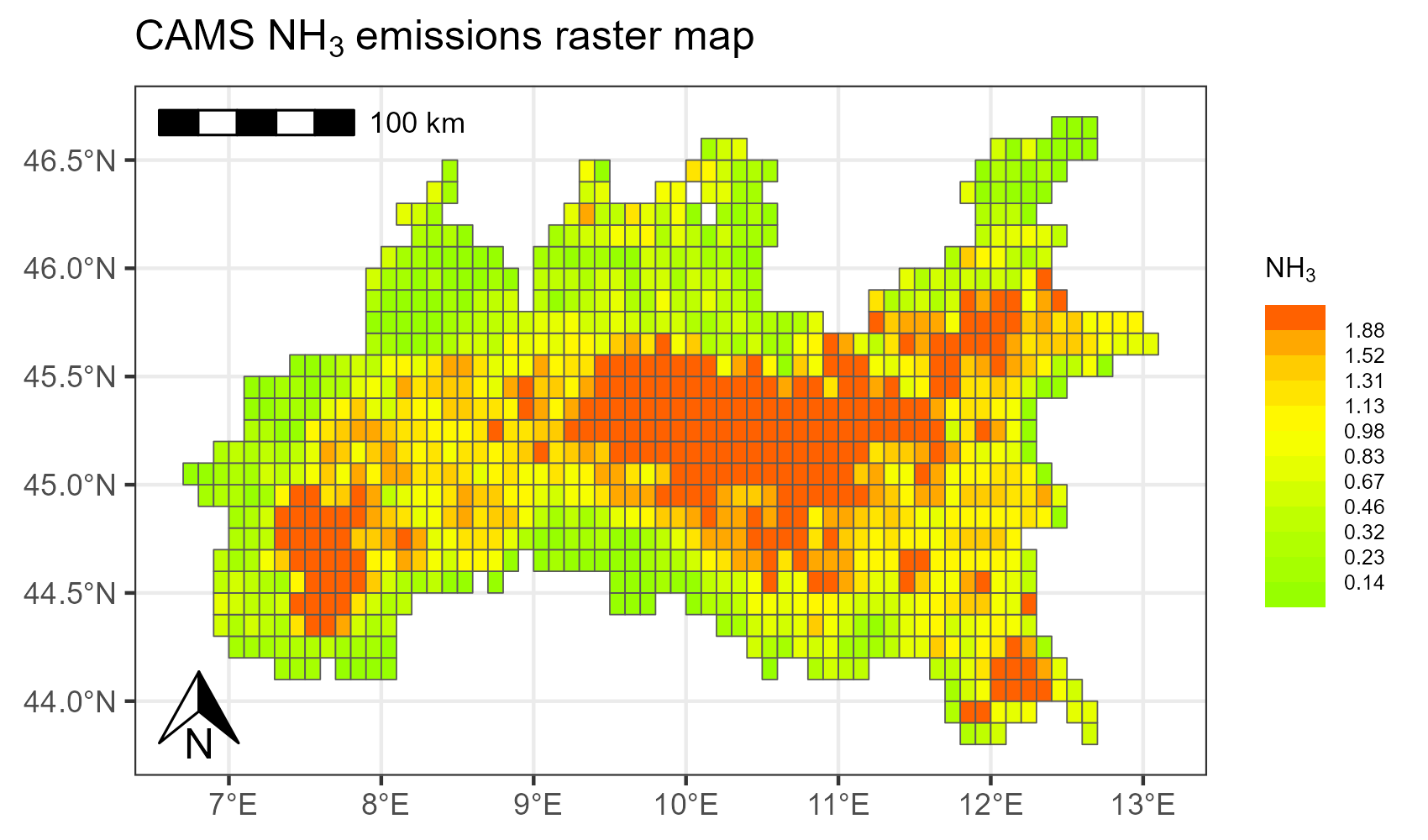}
\caption{Gridded map of the ammonia $NH_3$ emissions in 2020 in the Po Valley from the CAMS database. Pixels are regular squares with $0.1^\circ \times 0.1^\circ$ length.}
\label{fig:emission}
\end{figure}

Figure \ref{fig:emission} displays the spatial distribution of NH$_3$ emissions provided by CAMS. The map shows that there are 1,023 measurement points in total within the region of interest, with the number of points per ASR ranging from a minimum of 1 to a maximum of 12, with an average number of points per ASR as large as 4.1.

\section{Direct and model-based small area estimation of the agricultural carbon footprint} \label{sec:methods}

To estimate the contribution of agricultural activities to GHG emissions at the ASR level in the Po Valley —represented here by the FADN-based agricultural CF —we adopt a SAE approach. This allows us to integrate survey data with census and satellite information within a consistent statistical framework. Specifically, we first compute direct estimates of the agricultural CF and then improve them using SAE models that explicitly account for the spatial structure of the data.

\subsection{Direct estimation} \label{sec:directestimation}
We assume that the population, denoted as \( U \), consists of \( N \) units, which are distributed across \( m \) disjoint small areas. In our case, the population comprises \( N = 235,360 \) farms operating in the Po Valley, and the areas correspond to the \( m = 222 \) ASRs with at least one farm sampled in the FADN survey. Of these farms, \( n = 2{,}791 \) farms are sampled according to the design described above.

We use the subscript $i$ below to indicate a given area, hence $n_i$ and $N_i$ denote the sample size and the population size in area $i$, respectively. Let $y_{ik}$ denote the variable of interest, i.e. the CF of farm $k$ in area $i$.
The Horvitz-Thompson (HT) estimator of the population total of the CF $\tau_i$ in area $i$ is given by 
\begin{equation}
\hat{\tau}_i=\sum_{k=1}^{n_{i}} \frac{y_{ik}}{\pi_{ik}}
=\sum_{k=1}^{n_{i}} y_{ik}w_{ik},
\label{eq:direct}
\end{equation}

where $\pi_{ik}$ is the probability of including unit $k$ in area $i$ in the sample (first order inclusion probability) and the unit weight satisfies 
\[
\pi_{ik}=\frac{1}{w_{ik}}.
\]

The sample is selected from the population using the sampling design described above, and the inclusion probabilities of each single unit are implicitly defined by the sampling stratification. As noted earlier, the sampling design was not representative at the ASR level; rather, it was conducted at the regional level.

Since we aim to estimate the agricultural CF at a finer spatial scale, an appropriate weighting scheme is essential for computing reliable small-area estimates. To achieve this, we post-stratified the sample using census data, which involved calculating weights for the sampled units after the sampling process, based on auxiliary information about the ASR.

More specifically, the 2020 Census provides information on the number of farms by economic size and technical specialization at the municipal level. Since each municipality belongs to exactly one ASR, we were able to aggregate the Census data at the ASR level.

Next, we calculated the joint frequencies of farm holdings based on economic size (small vs. large farms), technical specialization (crop specialized, livestock specialized, and mixed farming), and agrarian subregion. Using this information, we developed a new proportional weighting scheme for each farm \( k \) in the FADN sample located in the \( i \)-th ASR, which is given by:
\[
\tilde{w}_{istk} = \frac{N_{ist}}{n_{ist}}
\]
where \( i = 1, \ldots, m\,(222) \) represents the considered ASR, \( s = 1, 2 \) indicates the economic size, and \( t = 1, 2, 3 \) denotes the farming typology.

Once the new weights have been obtained, a direct estimate of the population total for each ASR can be computed using equation (\ref{eq:direct}), where the original sampling weights $w$ are replaced by the calibrated weights $\tilde{w}$, namely:
\begin{equation}
\tilde{\tau}_i=\sum_{s=1}^{2} \sum_{t=1}^{3} \sum_{k=1}^{n_{ist}} y_{istk} \tilde{w}_{istk}
\label{eq:HTE}
\end{equation}
with estimated variance \cite{MoralesSAE} given by:

\begin{equation}
 \widehat{\mathrm{Var}}
 (\tilde{\tau_i})=\sum^2_{s=1}\sum^3_{t=1} 
 \sum_{k = 1}^ {n_{ist}} \tilde{w}_{istk}(\tilde{w}_{istk}-1)y_{istk}^2.
\label{eq:HTvar}
\end{equation}

\subsection{The Spatial Fay-Herriot model for log-Normal responses} \label{sec:SAE}
To obtain small area estimates of the CF for the ASRs in the region of interest, the Fay-Herriot method \cite{FH79} can be applied. 

In what follows, we assume that the population total $\tau$ of CF is related to the average CF $\mu$ through the relationship $\tau = N \times \mu$, where $N$ denotes the size of the population of interest.

To smooth the expected value of CF we adopt a model-based approach in which $\mu$ is expressed as a function of a set of area-specific covariates. Once the expected value has been obtained, the CF total for a given area is derived using the above relationship. To this end, we employ the area-level model proposed by Fay and Herriot, hereinafter referred to as the FH model. However, exploratory analysis suggests that the CF area means can hardly be considered normally distributed (see the right panel of Figure~\ref{fig:HTmap}), and that a logarithmic transformation of the data would be more appropriate instead. For this reason, the FH approach was applied to the log-transformed direct estimates of the area means $\mu_i$.

The FH model is based on two stages: the sampling model (first stage)
\[ 
\log \tilde{\mu_i} = \log \mu_i + \epsilon_i 
\]
and the linking model (second stage)
\[
\log \mu_i = x'_i \beta + u_i
\]
where $x_i$ and $\beta$ denote the $k \times 1$ vectors of area level covariates and regression parameters respectively. The sampling errors $\epsilon_i$ are assumed to be $0$-mean independent homoscedastic normally distributed random variables. The random effects $u_i$ are also assumed to be homoscedastic and independently normally distributed. For further details, we refer to \cite{RaoMol2015} or to \cite{MoralesSAE}. The combination of both models leads to an area-level linear mixed model.

Given the expected spatial dependency across geographically close regions, a random effect of simultaneous autoregression (SAR) \cite{WHITTLE54} was incorporated to account for potential interactions between neighboring areas \cite{PratSal2008}. In matrix form, the model writes:
\[
\mathbf{\log \tilde{\mu}}=\mathbf{X}\beta+ u + \epsilon,
\]
where $\mathbf{\log \tilde{\mu}}$ is the column vector of the 222 direct estimates of the mean ASR CF on the log scale,
$\mathbf{X}$ is a $222 \times k$ design matrix that includes the values of $k$ measured covariates described in Section \ref{sec:data} and incorporated into the model as described in Section \ref{sec:results}, $\beta$ is the corresponding $k$-vector of regression coefficients and 
$u$ is a column vector of error terms. In the SAR model, it is assumed that
\[
u = \rho \mathbf{W} u + v,
\]
where $\rho$ is the parameter representing the strength of spatial dependence, $v$ is a vector of i.i.d.\ Gaussian components with mean zero and variance $\sigma^2_v$, and $\mathbf{W}$ is a row-standardized $222 \times 222$ spatial contiguity matrix. The elements of $\mathbf{W}$ are defined as
\[
w_{ij} = \frac{\mathrm{1}(i \sim j)}{\# \partial(i)},
\]
where $i \sim j$ indicates that region $i$ is adjacent to region $j$, $\partial(i)$ denotes the set of neighbors of region $i$, and $\#$ represents the cardinality of this set. Hence, this can be equivalently rewritten as $u = (\mathbf{I} - \rho \mathbf{W})^{-1} v$, where $\mathbf{I}$ is the $222 \times 222$ identity matrix. Therefore, the model can be reformulated as:
\begin{equation}
\mathbf{\log \tilde{\mu}} = \mathbf{X} \beta + (\mathbf{I}-\rho  \mathbf{W})^{-1} v + \epsilon.
\label{eq:sarmod}
\end{equation}

It is possible to create higher order SAR models where the spatial term incorporates different neighborhood structure in the form $\rho_1W_1 + \rho_2W_2 + \cdots+ \rho_k W_k$ with $W_b$ specifying neighborhoods of different order \cite{Haining90}. However, most SAR applications consider a single spatial parameter, since increasing the number of spatial parameters complicates the estimation process.
Estimation has been carried out via Restricted Maximum Likelihood (REML) using the R package \texttt{emdi} \cite{emdipackage, emdipackageFH}. This model will be named SFH (Spatial Fay-Herriot) in the rest of this paper.

The model enables smoothing of direct CF estimates for a given ASR $i$ by incorporating a set of covariates. In particular, it allows for the computation of the Empirical Best Linear Unbiased Predictor (EBLUP), which effectively combines direct survey-based estimates with model-based predictions.\\
Defining $\theta=(\sigma^2_v, \rho)$, the variance-covariance matrix of $u$ is given by:
\[
\mathrm{Var}(u)=\sigma^2_v[(\mathbf{I}-\rho \mathbf{W})(\mathbf{I}-\rho \mathbf{W})']^{-1}:=\mathbf{G}(\theta).
\]
Consequently, the variance of $\tilde{\mu}$ is:
\[
\mathrm{Var}(\mathbf{\log \tilde{\mu}})=\mathbf{G}(\theta)+\mathbf{V}_\epsilon:=\mathbf{V}(\theta),
\]

where $\mathbf{V}_\epsilon$ denotes a diagonal matrix whose $i$-th diagonal element corresponds to the variance $\sigma^2_{\epsilon_i}$ of the HT estimate on the logarithmic scale in the $i$-th ASR, estimated as \cite{bishop2007discrete}:
\[
\hat{\sigma}^2_{\epsilon_i}=
\widehat{
   \mathrm{Var}}(\log \tilde{\mu_i})=
\frac{\widehat{\mathrm{Var}}(\tilde{\tau_i})}{\tilde{\tau}_i^2}.
\]

According to the SFH model specification, the BLUP takes the following form:

\[
\widehat{\log\mu}_i=x'_i\hat{\beta}(\theta)+I'_i\mathbf{G}(\theta)\mathbf{V}^{-1}(\theta)[\mathbf{\log \tilde{\mu}}-\mathbf{X}\hat{\beta}(\theta)]
\]

where $\hat{\beta}(\theta)=(\mathbf{X}'\mathbf{V}^{-1}(\theta)\mathbf{X})^{-1}\mathbf{X}'\mathbf{V}^{-1}(\theta)\mathbf{\log \tilde{\mu}}$ and $I_i$ is the vector with 1 in the $i$-th position and 0 elsewhere. Spatial EBLUP can be obtained via the plug-in of a consistent estimator $\hat{\theta}=(\hat{\sigma}^2_v,\hat{\rho})$ for $\theta$, which in our case corresponds to the REML estimator \cite{MoralesSAE}.

The predicted mean value for area $i$ in its original scale, that is $\hat{\mu}_i$, is then obtained using the back-transformation formula proposed by \cite{slud2006mean}:
\begin{equation} 
    \hat{\mu}_i = \exp\left( \widehat{\log{\mu}_i} + \frac{
    \mathrm{MSE}\left( \widehat{\log{\mu}_i} \right)}{2} \right), 
\label{eq:EBLUP}
\end{equation}
where $\mathrm{MSE}\left( \widehat{\log{\mu}_i} \right)$ is the MSE of the SFH model prediction on the log scale for area $i$. Notice that, the quantity $\mathrm{MSE}\left( \widehat{\log{\mu}_i} \right)$ can be either computed using the analytical approximation provided by \cite{singh2005spatio} or a parametric bootstrap approach \cite{MoralesSAE}. In what follows, both methods will be considered.

\section{Up-scaling spatial covariates to the ASR level} \label{sec:upscale}
As mentioned in the previous sections, we aim at estimating the CF at the ASR level using a number of potential predictors. 
The proposed methodology builds on the SAE framework, specifically employing area-level models such as the SFH model. These models are designed for situations where the explanatory variables are associated with small domains (i.e., subregions), and they provide estimates directly for these domains.
In our case, however, these auxiliary variables are all collected at a spatial support that differs from the one, the agriculture sub-regions, that is the pertinent one for the problem at hand.  

The information provided by the Italian National Census of Agriculture and the National Veterinary Registry Office, described in Section \ref{sec:data}, is available at the municipality level (i.e., LAU — Local Administrative Units, second level according to the Eurostat nomenclature of administrative units). However, each municipality belongs to only one ASR; therefore, measures at the ASR level can be easily obtained by simply aggregating the data across municipalities within each subregion. As noted in Section \ref{sec:data}, CAMS data are instead provided on a fine spatial grid. Consequently, for the area-level model, it was necessary to realign these data to the ASR level in order to incorporate them as covariates.

The need to integrate data collected on different spatial supports in statistical analysis is known as the Change of Support Problem (COSP), as thoroughly described by \cite{gotway_young_2002}.
When data need to be up-scaled, block kriging (BK) represents a natural choice for accomplishing this task \cite{dowd2024many, belmonte2025approach}.

Let \( Y = \{Y(s), \, s \in D\} \) be a real-valued spatial random field, and let \( V \subset D \) be a subregion of interest of size $|V|$. The spatial average of \( Y \) over \( V \) is defined as 
\[
Y_V =|V|^{-1} \int_V Y(s) \, ds.
\]

Ordinary BK provides the Best Linear Unbiased Estimator (BLUE) of the block average using a linear predictor of the form
\[
\hat{Y}_V = \sum_{i=1}^k \alpha_i Y(s_i),
\]
where \( Y(s_1), \ldots, Y(s_k) \) are the observed values of \( Y \) at sampling locations \( s_1, \ldots, s_k \), and \( \alpha_1, \ldots, \alpha_k \) are weights satisfying the constraint
$\sum_{i=1}^k \alpha_i = 1$. In local BK, only a subset of the sampling locations - typically the ones closest to the subregion of interest - is used in the linear combination to predict the block average.

Determining the kriging weights requires minimizing the mean squared error of the prediction, that is \( E(Y_V - \hat{Y}_V)^2 \), under the constraint mentioned above. This involves calculating the block variogram between each sample point and the block, i.e., 
$\frac{1}{|V|} \int_V \gamma(s_i - s) \, ds$, and the variogram within the block, i.e., $\frac{1}{|V|^2} \int_V \int_V \gamma(x - s) \, dx \, ds$, where $\gamma(\cdot)$ is the semi-variogram function of the random field.
In practice, these integrals are approximated by weighted sums based on appropriate quadrature schemes. A detailed description of BK can be found in \cite{chiles2012geostatistics}. 

\begin{figure}
\centering
\includegraphics[width=0.45\linewidth]{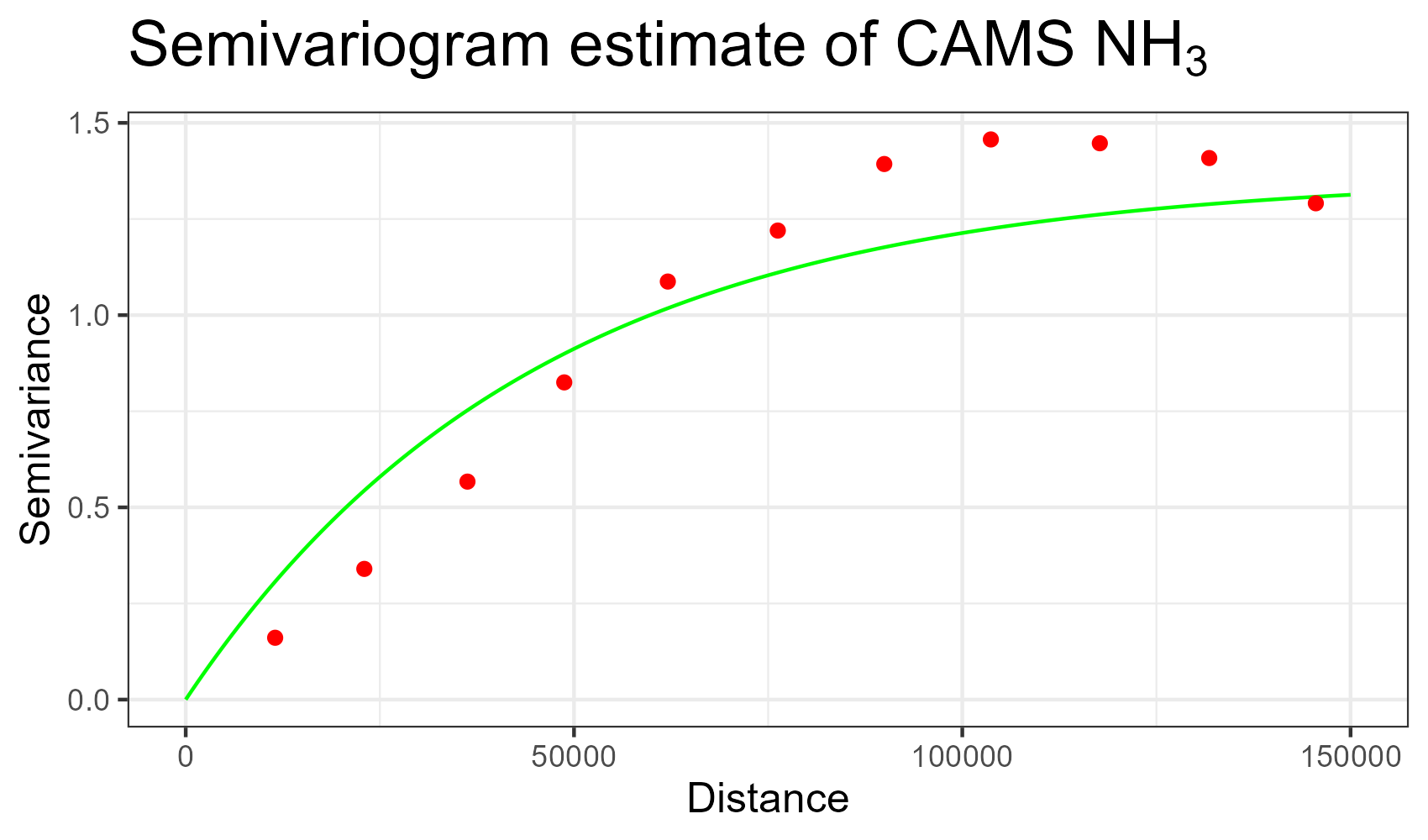}
\includegraphics[width=0.45\linewidth]{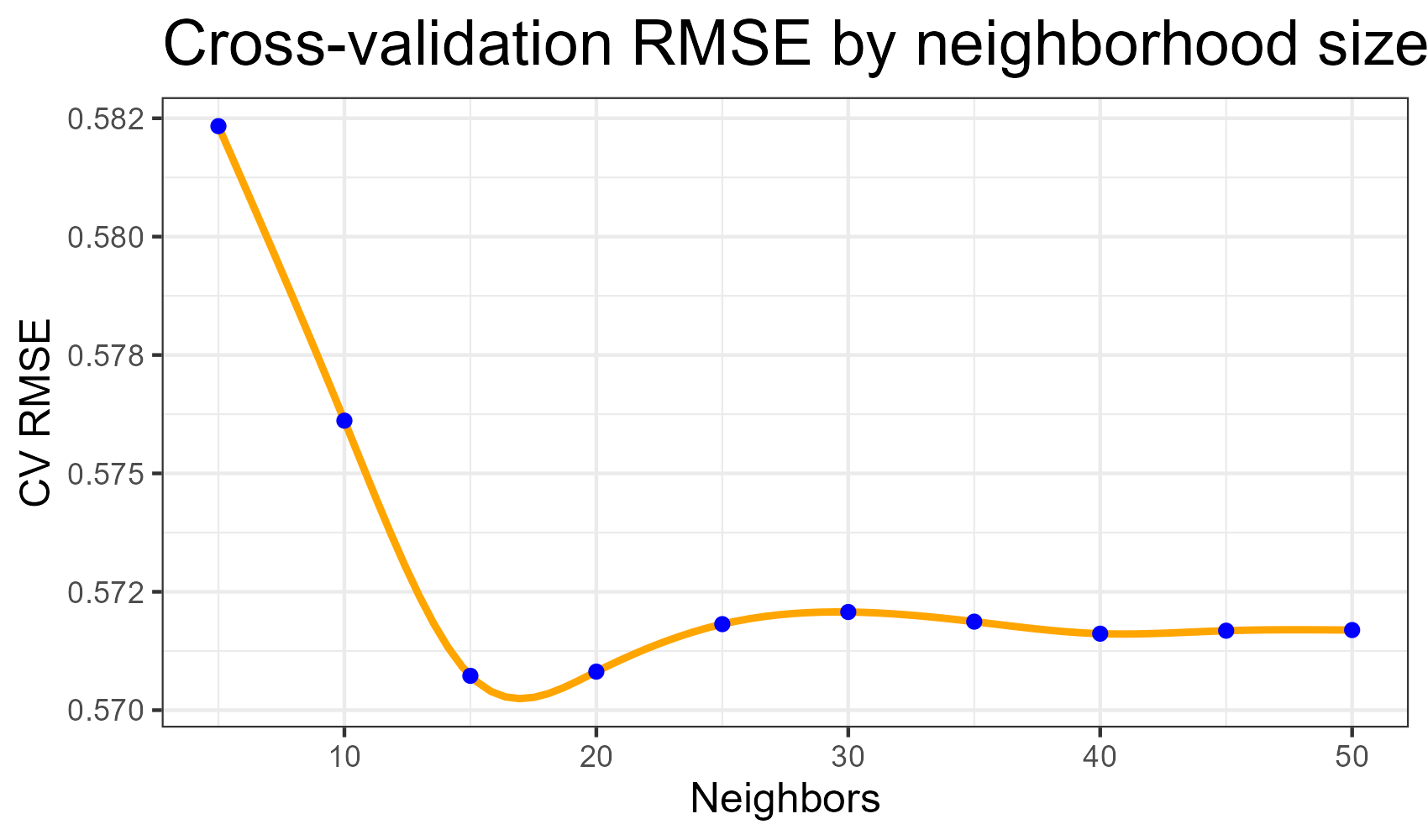}
\includegraphics[width=0.45\linewidth]{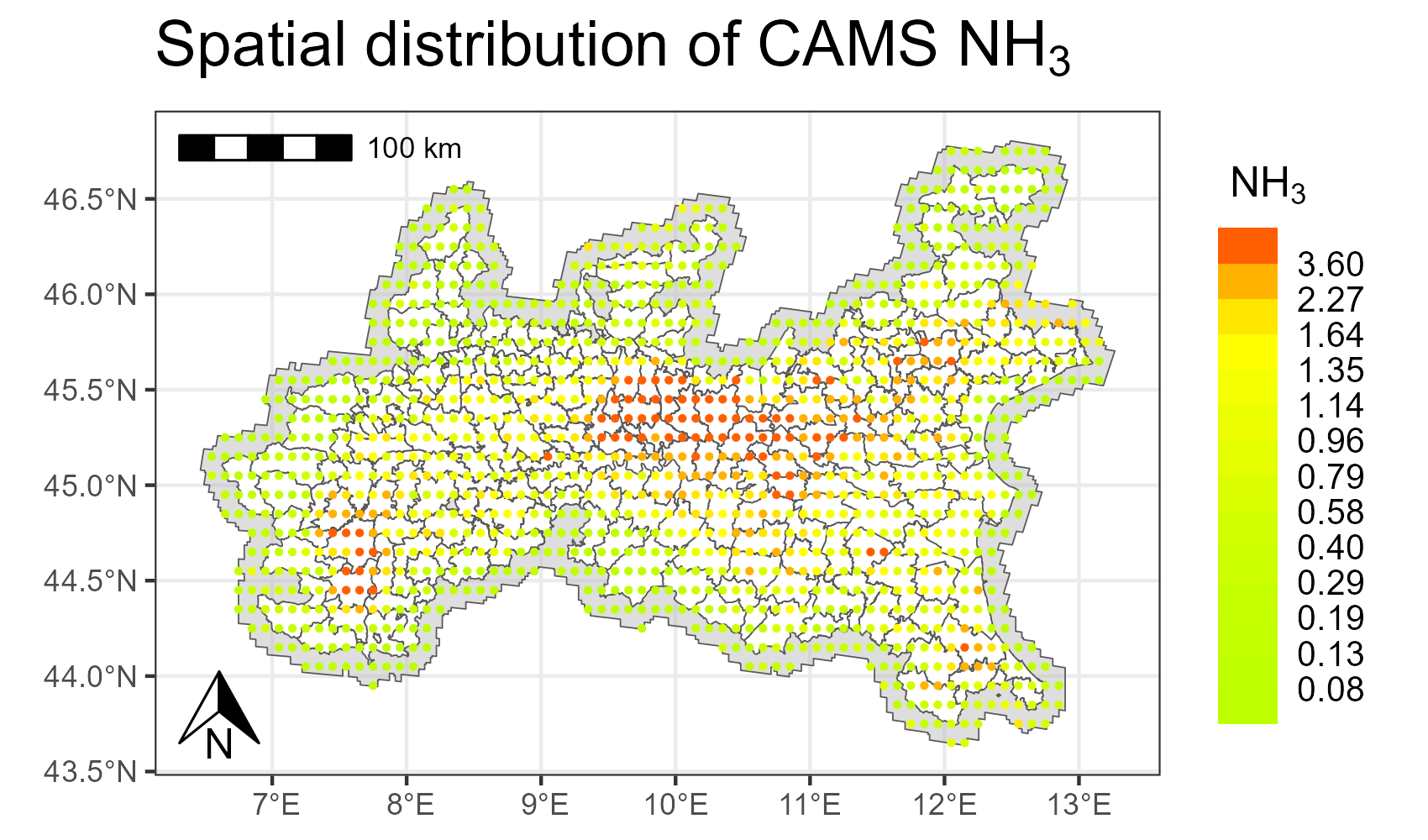}
\includegraphics[width=0.45\linewidth]{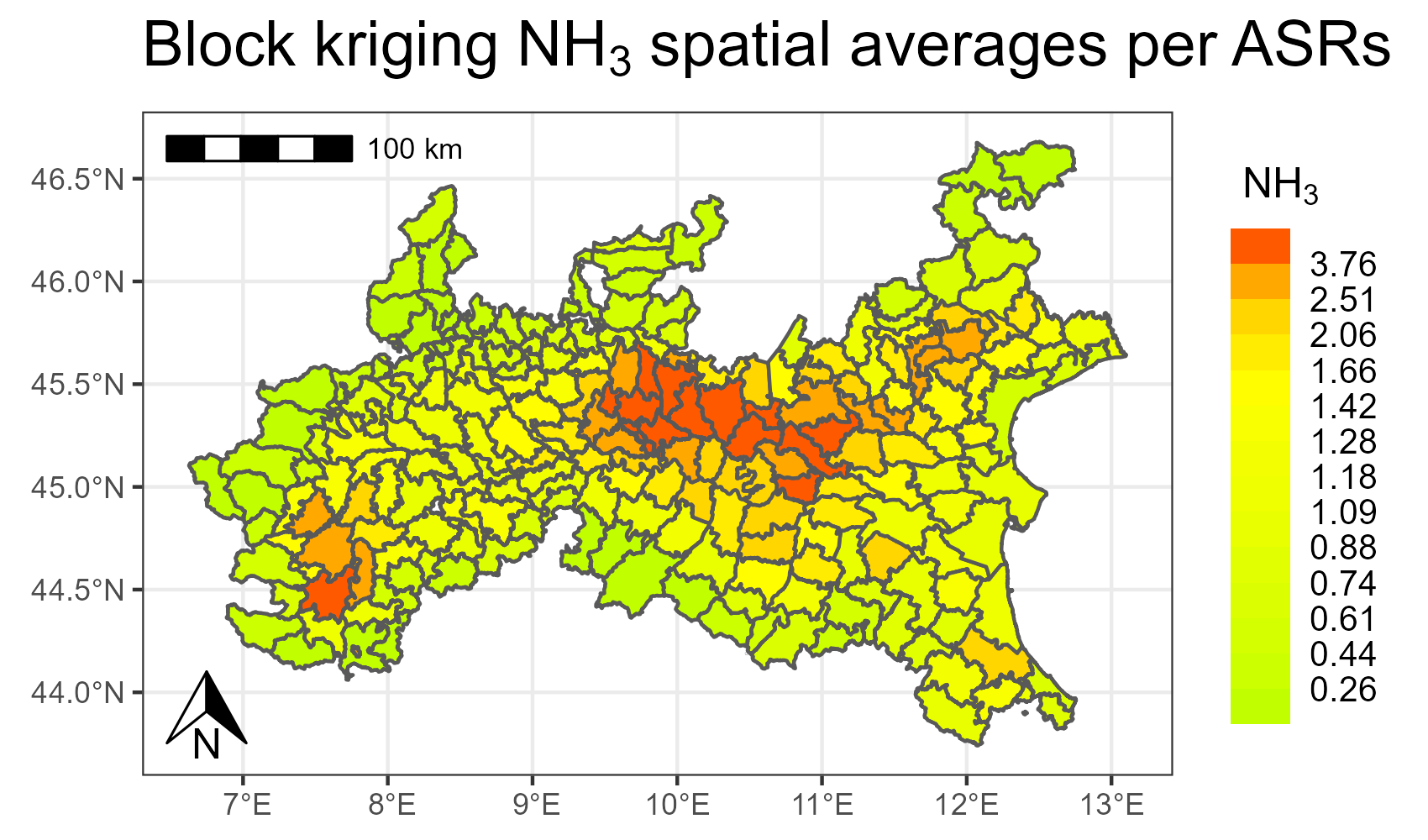}
\caption{Top-left panel: (a) Estimated semi-variogram for the NH$_3$ emissions on the $0.1^\circ \times 0.1^\circ$ regular grid. Top-right panel: (b) Average 5-fold cross-validation RMSE by neighborhood size. Bottom-left panel: (c) CAMS data for NH$_3$ emissions in the 222 considered ASRs with a $10$km buffer. Bottom-right panel: (d) Estimated average NH$_3$ emissions per ASR by block kriging.}
\label{fig:BloKriging}
\end{figure}

In our case, \( V \) represents one of the 222 ASRs with at least one farm included in the FADN sample, while \( k = 1,259 \) corresponds to the number of values provided by CAMS at the grid points within the area of interest extended with a buffer of $10$km to protect towards potential boundary effects, and displayed in Figure~\ref{fig:BloKriging}-c. To calculate BK values, the Matérn variogram model was fitted to the data using ordinary least squares. Figure \ref{fig:BloKriging}-a shows the estimated variogram. The Matérn model not only outperformed other parametric variogram models in a cross-validation procedure but is also coherent on most applied studies on the spatial modeling of GHG emissions (e.g.,\cite{western2020bayesian}).

We then applied local ordinary BK, with a kriging neighborhood of 15 points, to obtain spatial averages of emissions at the ASR level. The optimal neighborhood size was selected via \(5\)-fold cross-validation. Figure \ref{fig:BloKriging}-b displays the resulting cross-validation root mean squared error (RMSE) curve for different neighborhood sizes. Figure \ref{fig:BloKriging}-d displays the spatial averages at the ASR level obtained by BK. The \texttt{R} package \texttt{gstat} was used to implement the procedure \cite{pebesma2004gstat}.

\section{Double parametric bootstrap procedures for uncertainty propagation} \label{sec:DoubleBoot}
The upscaling procedure described above provides an estimate of the actual $NH_3$ emissions. Consequently, the additional variability arising from the estimation of emissions at the ASR level must be properly accounted for during the model estimation phase to ensure valid standard errors and confidence intervals. The introduction of an error component is not new in the SAE literature. For example, the popular model by \cite{ybarra2008small} considers the measurement error in the covariates, while recently \cite{battagliese2026downscaling} proposed a three-stage FH model where the additional stage accounts for the error in the direct estimator.

Relatively to the uncertainty arisen from the upscaling of satellite-based covariate, \cite{chi2022microestimates} proposed a naive approach based on a linear regression model on the absolute residuals but lacks in a formal description of the method; on the other hand, \cite{wieczorek2023design} proposed the use of conformal prediction but it requires the assumption of exchangeability which is not suitable for spatially dependent data. However, a fully satisfactory solution to address this issue has not been found yet \cite{tzavidis2025small}.

To this end, we propose a double resampling procedure that combines a parametric bootstrap with conditional stochastic simulations of the trajectory of a spatial random field. At each bootstrap iteration, data are sampled from the estimated model, where the covariate (average $NH_3$ emission) is obtained by block kriging (BK) of simulated data from a spatial random field at a given set of spatial locations. The simulations are carried out using a conditional approach generating the data from a Gaussian process with a variogram function defined by the Matérn model estimated on the actual dataset. Details on the simulation process are provided in Appendix \ref{App_B}.

The double parametric bootstrap procedure presented in Algorithm \ref{AlgDoubleBoot} allows to account for the additional variability induced by the upscaling procedure in the estimation of the standard errors of the model estimators and the MSEs of the EBLUPs. Whereas the standard errors of regression coefficients $\hat{\beta}$ and spatial autocorrelation $\hat{\rho}$ are simply obtained by the standard deviation between the estimates at different iterations, the MSEs of the EBLUP in ASR $i$ is computed as: 
\[
\mathrm{MSE}\left( \widehat{\log{\mu}_i} \right)=\frac{1}{B}\sum_{b=1}^B\left(\widehat{\log \mu^{*}_{i,b}}-\log \mu_{i,b}^{*}\right)^2
\]
where $\log \mu^{*}_{i,b}$ is the bootstrap mean simulated from the SFH linking model and $\widehat{\log \mu^{*}_{i,b}}$ is the spatial EBLUP for region $i$ at iteration $b$ (see details in Algorithm \ref{AlgDoubleBoot}).

\begin{algorithm}
\caption{Double parametric bootstrap for standard error and confidence interval computation}
\label{AlgDoubleBoot}
\begin{algorithmic}[]
	\State set $B$: number of bootstrap iterations
	\State \,\,\,\,\, \, $A$: an empty list of size $B$ 
	\State \,\,\,\,\, \, $b=1$
	\State Load the border of the region of interest $D$
    \While{$b \leq B$}
        \State Simulate 1259 spatial locations from the uniform distribution on $D$ 
        \State Simulate a trajectory of $NH_3$ at the simulated locations from
        \State \,\, a log-gaussian process with the variogram estimated on the data
        \State Calculate BK prediction at each ASR 
        \State Simulate the HT on the log-scale using the estimated SFH model
        \State Estimate the SFH model using the simulated HT estimates 
        \State \,\, and the BK predictions obtained from simulation; 
        \State Save the estimated parameters: $A[[b]]\leftarrow (\widehat{\beta}^*_b, 
        \widehat{\rho}^*_b)$
        \State $b=b+1$     
                \EndWhile
    \State \Return Standard error and percentile confidence intervals
\end{algorithmic}
\end{algorithm}

The bootstrap procedure described above can be extended to perform hypothesis testing on the model parameters. In particular, a Monte Carlo test can be implemented within the double parametric bootstrap scheme in Algorithm \ref{AlgDoubleBoot} to compute $p$-values for any parameter of interest and check whether it is statistically significant or not.

The model is first estimated on the actual data with the parameter of interest set to zero. Let us denoted this model as $\widehat{m}_{0,obs}$ below. The unrestricted model is also estimated on the actual data and denoted by $\widehat{m}_{1,obs}$. The observed log-likelihood ratio is then computed as $l_{\text{obs}} = \log L_1 - \log L_0$, where $L_j$ is the likelihood of model $\widehat{m}_{j,obs}$, $j = 0, 1$. The previous steps are repeated within each bootstrap iteration for $B$ times, where the data are simulated from the restricted model estimated using the observed data, i.e., $\widehat{m}_{0,\text{obs}}$. This allows one to estimate the distribution of the test statistic under the restricted model (i.e., under the null hypothesis). Hence, at the end, the $p$-value can be computed as
\[
p\text{-value} = \frac{1}{B + 1} \sum_{b = 0}^{B} I(l^*_b \geq l_{\text{obs}}),
\]
where $l^*_b$ denotes the bootstrap replicate of the log-likelihood ratio statistic and $l_{\text{obs}}$ is its observed counterpart, where we set $l_0 = l_{\text{obs}}$. The same procedure can be repeated for each parameter of interest in order to obtain the corresponding $p$-value for each of them.

The full procedure is summarized in Algorithm \ref{AlgDoubleBootPV}\footnote{The part in parentheses is unnecessary when testing the null hypothesis that the regression coefficient for average NH$_3$ emission equals zero.}.

\begin{algorithm}
\caption{Double parametric bootstrap for $p$-value computation}
\label{AlgDoubleBootPV}
\begin{algorithmic}[]
	\State set $B$: number of bootstrap iterations
	\State \,\,\,\,\, \, $A$: an empty vector of size $B+1$ 
	\State \,\,\,\,\, \, $b=1$ 
	\State Load the border of the region of interest $D$
    \State Estimate the FH unrestricted and restricted model 
    $\widehat{m}_{0,obs}$ and $\widehat{m}_{1,obs}$ 
    \State Calculate $l_{\text{obs}} = \log L_1 - \log L_0$
    \State $A[b]=l_{\text{obs}}$
    \While{$b \leq B+1$}
        \State Simulate 1259 spatial locations from the uniform distribution on $D$ 
        \State Simulate a trajectory of $NH_3$ at the simulated locations of a 
        \State \,\, log-gaussian process with the variogram  estimated on the data    
        \State Calculate BK prediction at each ASR 
        \State Simulate the HT on the log-scale using $\widehat{m}_{0,obs}$
        \State Estimate the restricted model $\widehat{m}^*_{0,b}$ on the simulated HT estimates
          \State \,\, (and the BK predictions obtained from simulation)   
          \State Estimate the unrestricted model $\widehat{m}^*_{1,b}$ on the simulated HT estimates
          \State \,\, and the BK predictions obtained from simulation 
        \State Calculate $l^*_{b} = \log L^*_{1,b} - \log L^*_{0,b}$ 
        \State  $A[b]\leftarrow l^*_{b}$
        \State $b=b+1$
               \EndWhile
    \State \Return $p\text{-value} = \frac{1}{length(A)} \sum_{b = 1}^{B+1} A[b] \geq A[1]$
\end{algorithmic}
\end{algorithm}

\section{Empirical results} \label{sec:results}
This section presents the results obtained by applying the various approaches described above to the carbon footprint (CF) data introduced in Section \ref{sec:data}.

\subsection{Direct estimation of the Carbon Footprint}
\begin{figure}
\centering
\includegraphics[width=5.1cm]{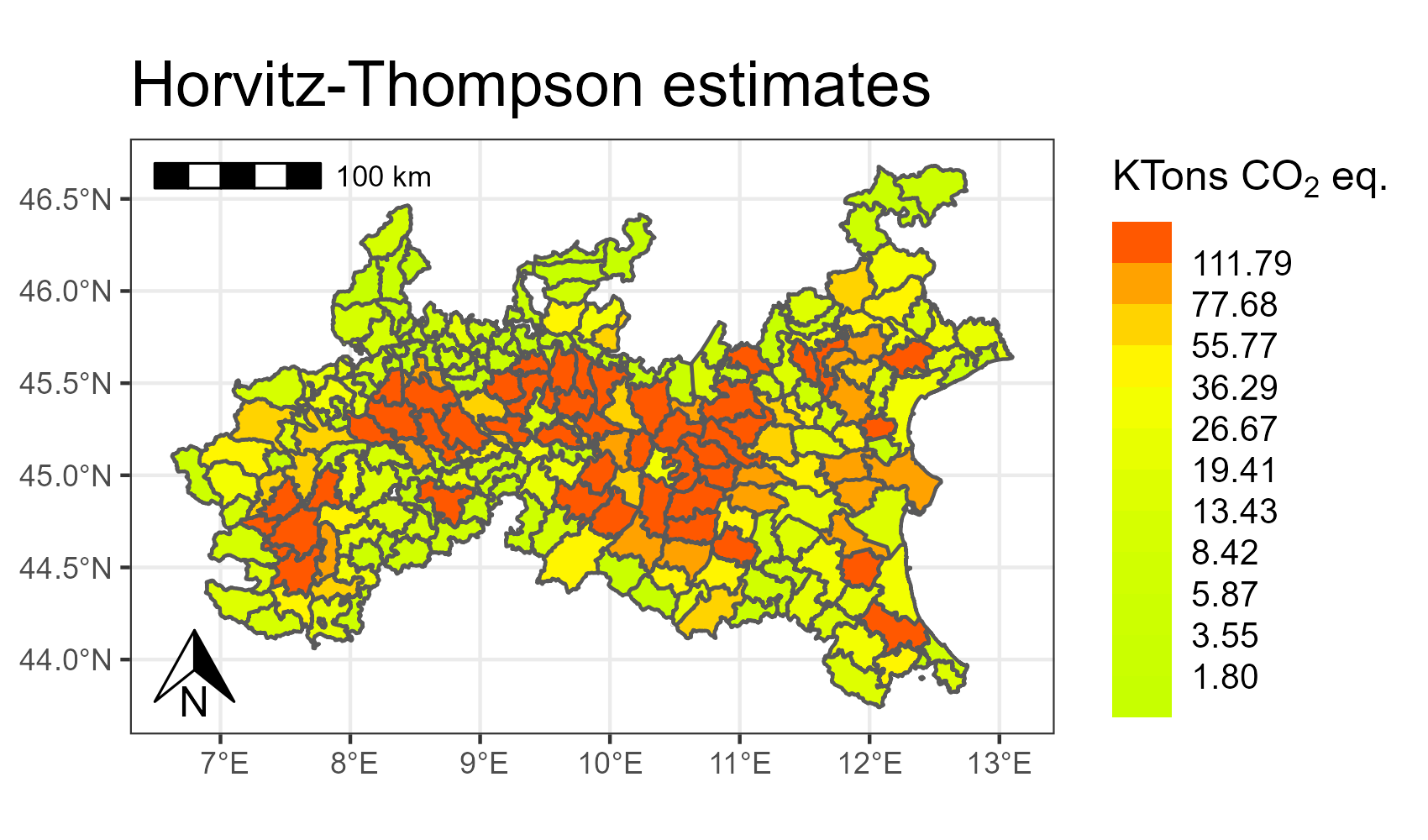}
\includegraphics[width=5.1cm]{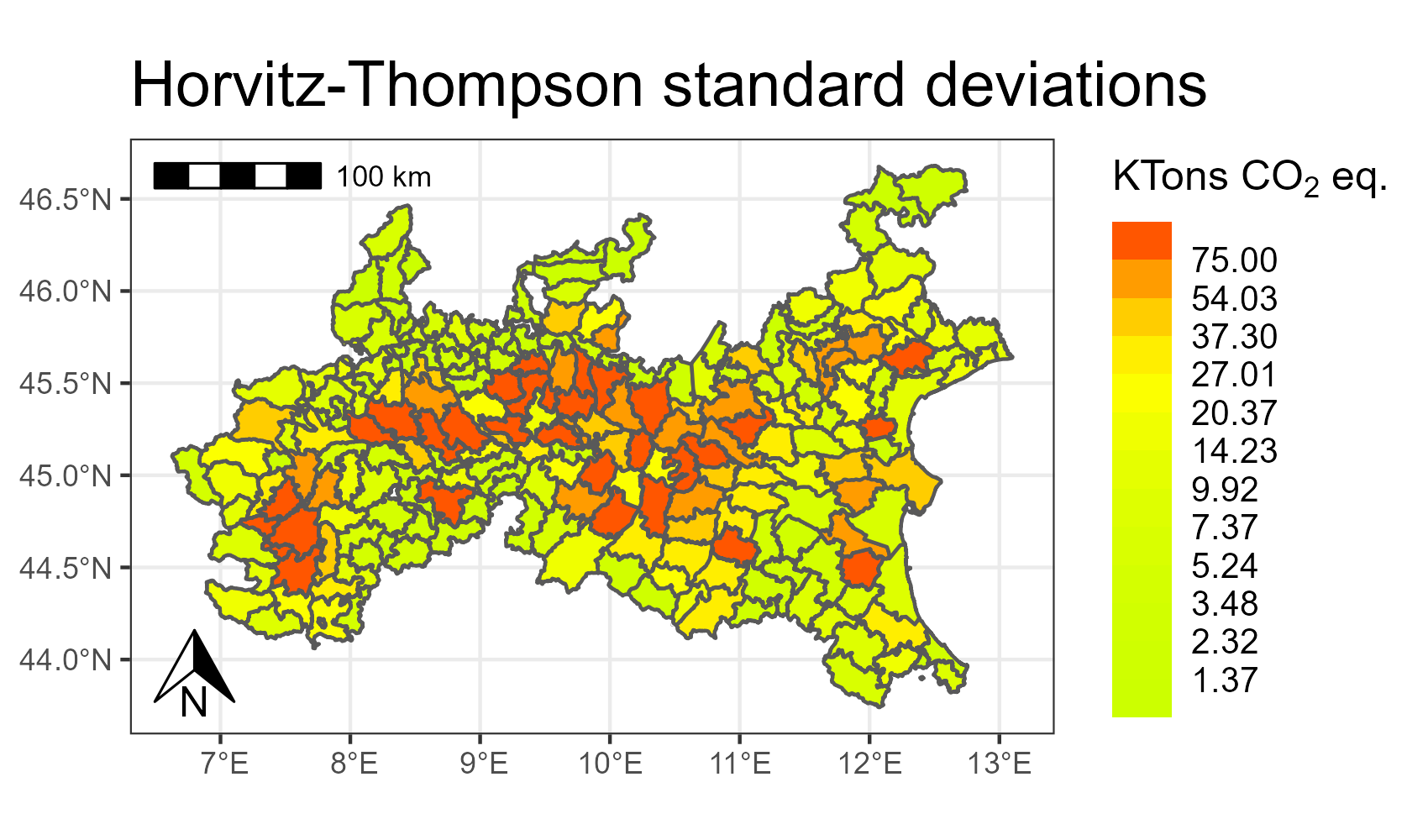}
\includegraphics[width=4.7cm]{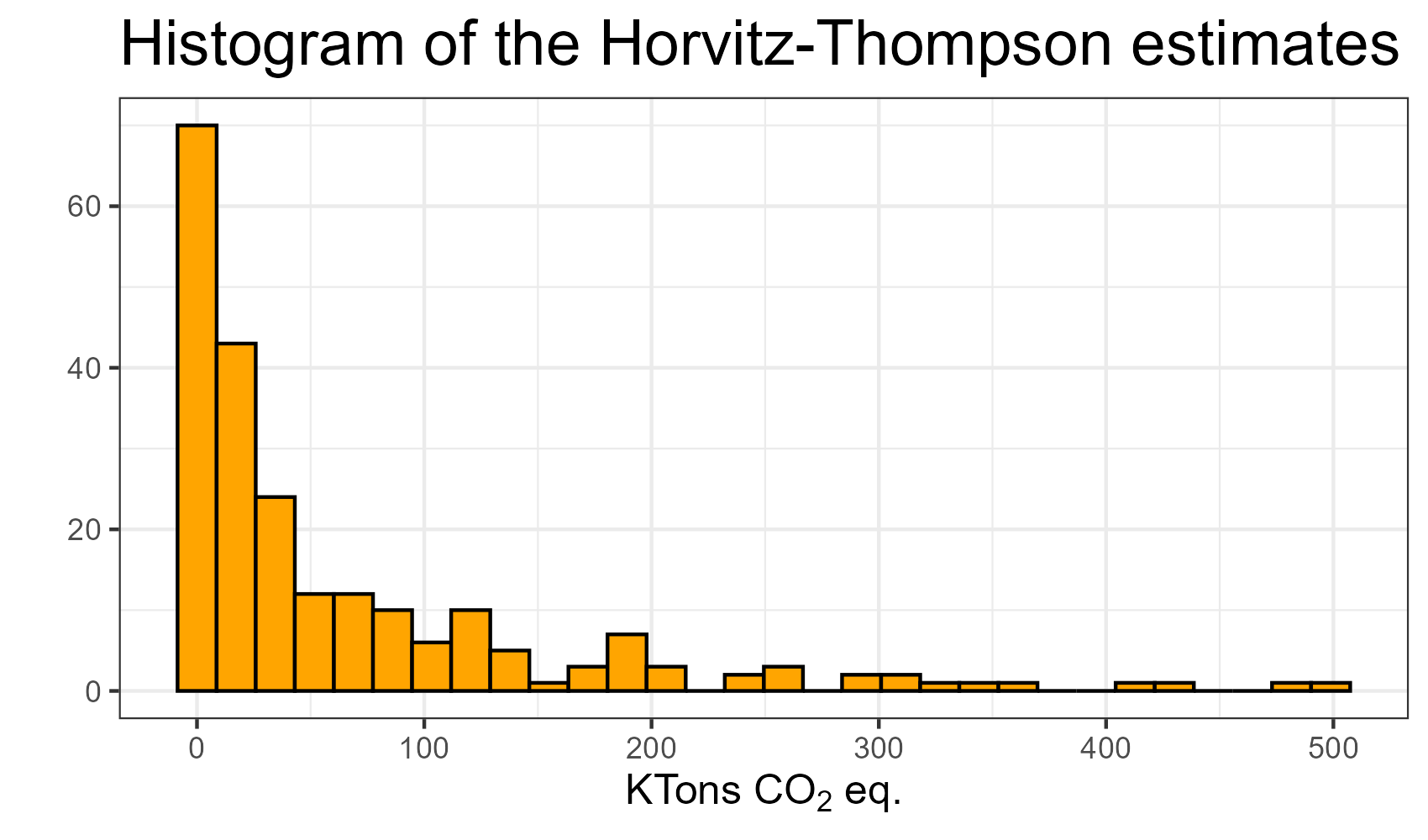}
\caption{Left panel: (a) Horvitz–Thompson direct estimates of the agricultural carbon footprint per ASR (KTons CO$_2$ eq.). Central panel: (b) empirical standard deviations of the Horvitz–Thompson direct estimates of the agricultural carbon footprint per ASR. Left panel: (c) histogram of the Horvitz–Thompson direct estimates of the agricultural carbon footprint per ASR.}
\label{fig:HTmap}
\end{figure}

Figure \ref{fig:HTmap} displays the CF estimates for each ASR in the Po Valley, as obtained using the HT estimator. Table \ref{tab:HTsummarystats} presents the main descriptive statistics for both the HT estimates and their associated estimated variances across the ASRs.

As mentioned above, the FADN sample is not designed to be representative at the ASR level, as the stratification follows a broader and less granular structure (i.e., regional stratification). 
For this reason, the post-stratification strategy described in Section \ref{sec:directestimation} was adopted.

The introduction of spatially representative weights has significantly improved the resulting map compared to using the original weighting scheme (results not reported here), yielding a smoother distribution and eliminating anomalous values in the mountainous areas of the north.

Despite their desirable statistical properties, however, direct estimates have been found to only partially capture the spatial distribution of the agro-CF and suffer from some instability. In particular, HT estimates of CF tend to fluctuate considerably, sometimes yielding unusually high or low values. Notably, certain regions near the Alps - generally considered to be relatively unpolluted - show unexpectedly high levels of CF.

\begin{table}
 \caption{Descriptive statistics of the HT estimates of CF} \label{tab:HTsummarystats}
 \centering
 \small
\begin{tabular}{@{}lccccccc@{}}
\toprule
    & Min & 1st Q & Median & Mean & 3rd Q & Max & Std. Dev. \\ 
\midrule
    CF: HT estimate & 0.05 & 6.18 & 24.14 & 65.35 & 81.70 & 498.80 & 93.24 \\
    CF: HT standard deviation & 0.05 & 3.79 & 11.92 & 33.74 & 42.24 & 329.38 & 50.49 \\
\bottomrule
\end{tabular}
\end{table}

In addition, the HT estimator has several limitations. First, it relies on the assumption that the sampled farms are fully representative of the population and does not incorporate any auxiliary information. In extreme cases, such as when no sampled farms are present in a given ASR despite the existence of farms in the actual population, the CF estimate will be  not available. Second, sample sparsity at the ASR level can lead to unreliable results, including implausible negative estimates of the CF.

For these reasons, in the next section we introduce a model-based estimate of the CF at the ASR level.

\subsection{Model-based estimation of the Carbon Footprint using satellite and census data}

We estimated the SFH model described in Section \ref{sec:SAE} using two alternative specifications. The first model includes the full set of variables described in Section \ref{sec:data}, excluding NH$_3$ emissions. This model will be denoted by \textit{Model 1} in the rest of the paper. The other, denoted by \textit{Model 2} below, includes only the emissions NH$_3$ obtained from satellite data in the linear predictor.

Recall that CAMS data on NH$_3$ emissions were upscaled from the regular grid scale to the ASR scale, hereafter denoted by \( NH3_i \), using the BK strategy described in Section~\ref{sec:upscale}.

The empirical specification of the SFH model in Equation \ref{eq:sarmod}, that is, Model 2, writes as
\[
\log \tilde{\mu}_i= \alpha_0 + \alpha_1 NH3_i + I'_i (\mathbf{I}-\rho \mathbf{W})^{-1} v + \epsilon_i
\]
where $\log \tilde{\mu}_i$ is the HT estimator of the mean CF and $NH3_i$ is the BK spatial average of NH$_3$ in ASR $i$, respectively. Then, the model-based estimate for the total CF is obtained as $\hat{\tau}_i=N_i\hat{\mu}_i$ and $\hat{\mu}_i$ is the EBLUP defined in Equation \ref{eq:EBLUP}.

Table \ref{tab:coefficientsSFH} shows the estimates of the regression coefficients for the two SFH models. Most of the estimated coefficients are positive. This result is consistent with expectations, that is the agricultural CF is generally expected to increase with the utilized agricultural area, the number of cattle units, and the NH$_3$ emissions. Two remarkable exceptions consist in the working days and the number of swine. In the latter case the estimate is not statistically significant, which may seem counterintuitive but it can be explained with the high correlation of these two variables and the utilized agricultural area and the bovines, respectively. Furthermore, all variables, except for the average altitude and the total swine heads, are found to be statistically significant.

\begin{table}
\caption{Estimated regression coefficient estimates for the Spatial Fay-Herriot model with SAR random effects. Standard errors are provided under parentheses. The covariates included in the linear predictor of both models have been standardized.}
\label{tab:coefficientsSFH}
\centering
\small
\begin{tabular}{@{}lcc@{}}

\toprule
\textbf{Variable} 
& \textbf{Model 1} & \textbf{Model 2} \\
\midrule
\multirow{2}{*}{Intercept} 
  & $3.454$ $^{***}$ & $3.451$ $^{***}$  \\
  & \footnotesize{($0.207$)} & \footnotesize{($0.226$)}  \\
\multirow{2}{*}{Utilized agricultural area} 
  & $0.587$ $^{***}$ & -- \\
  & \footnotesize{($0.138$)} & \\
\multirow{2}{*}{Standard working days} 
  & $-0.342$ $^{**}$ & -- \\
  & \footnotesize{($0.117$)} & \\
\multirow{2}{*}{Total bovine heads} 
  & $0.467$ $^{**}$ & --  \\
  & \footnotesize{($0.182$)} & \\
\multirow{2}{*}{Total swine heads} 
  & $-0.108$ & --  \\
  & \footnotesize{($0.180$)} &  \\
\multirow{2}{*}{Average altitude} 
  & $0.161$ & --  \\
  & \footnotesize{($0.141$)} &  \\
\multirow{2}{*}{NH$_3$ emissions} 
  & -- & $0.609$ $^{***}$ \\
  & & \footnotesize{($0.121$)} \\
 \multirow{2}{*}{Spatial autocorrelation $\rho$} 
   & $0.717$ & $0.733$ $^{***}$  \\
   & & \footnotesize{($0.069$)}  \\
\bottomrule
\end{tabular}
\end{table}

\begin{figure}
\centering
\includegraphics[width=5.1cm]{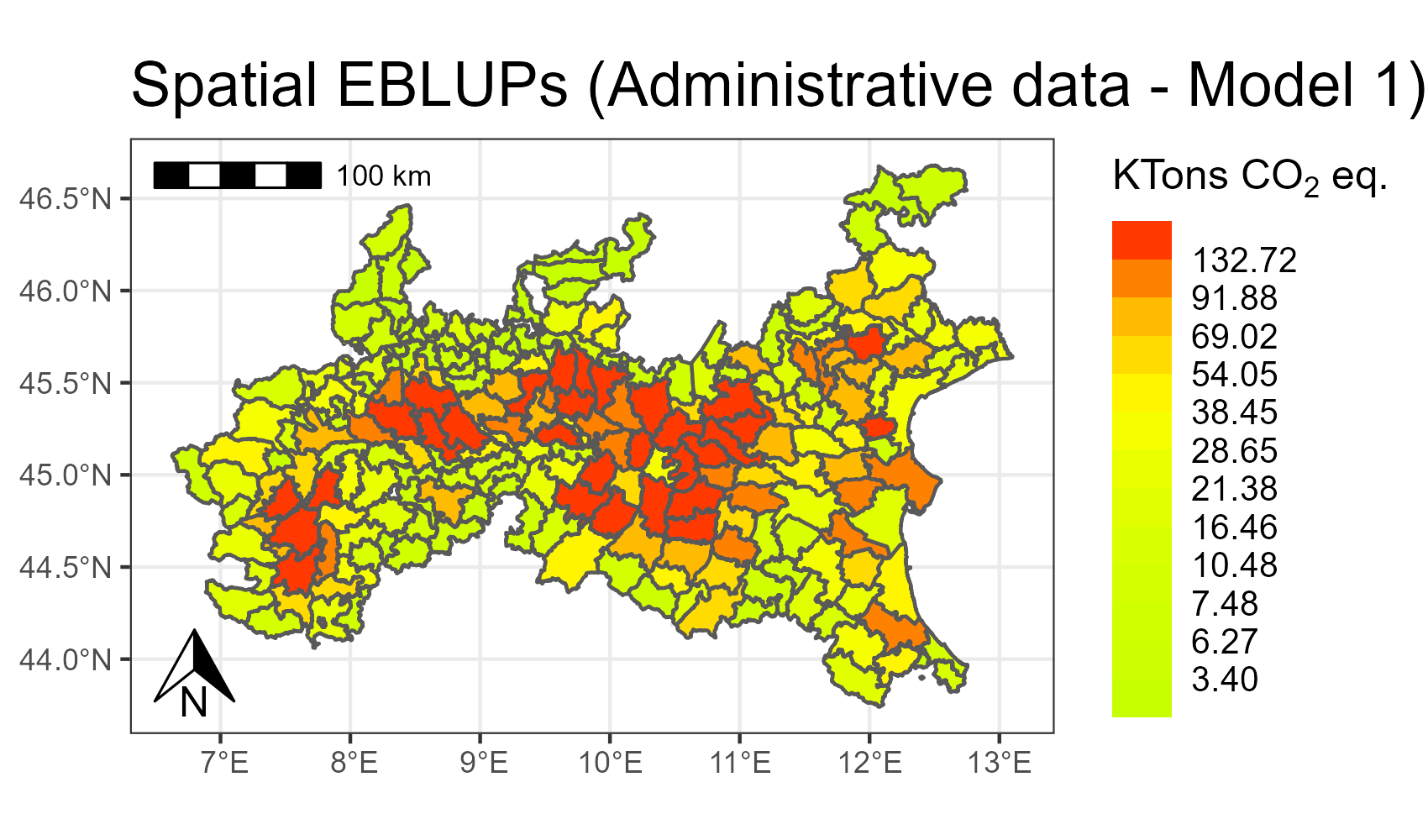}
\includegraphics[width=5.1cm]{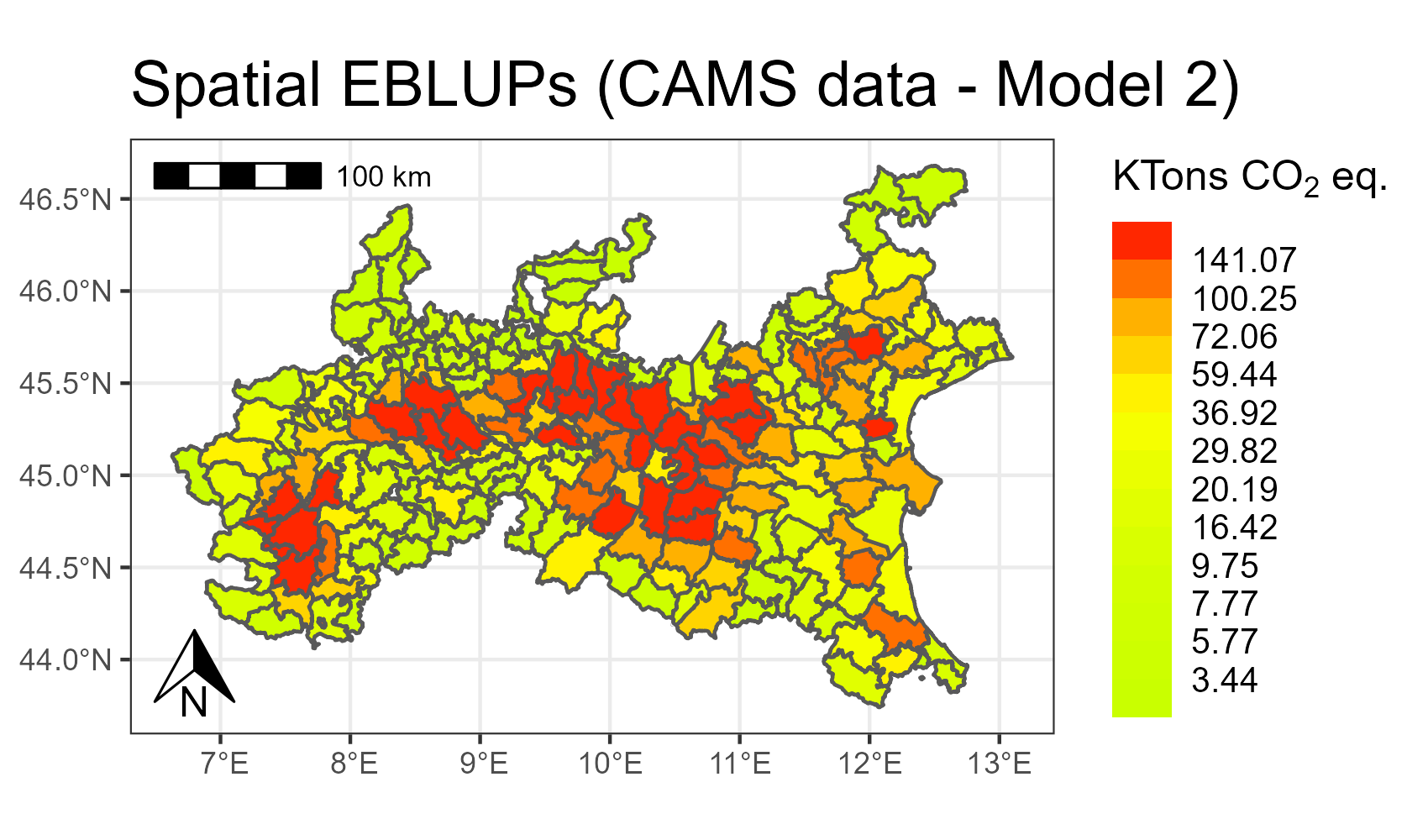}
\includegraphics[width=4.7cm]{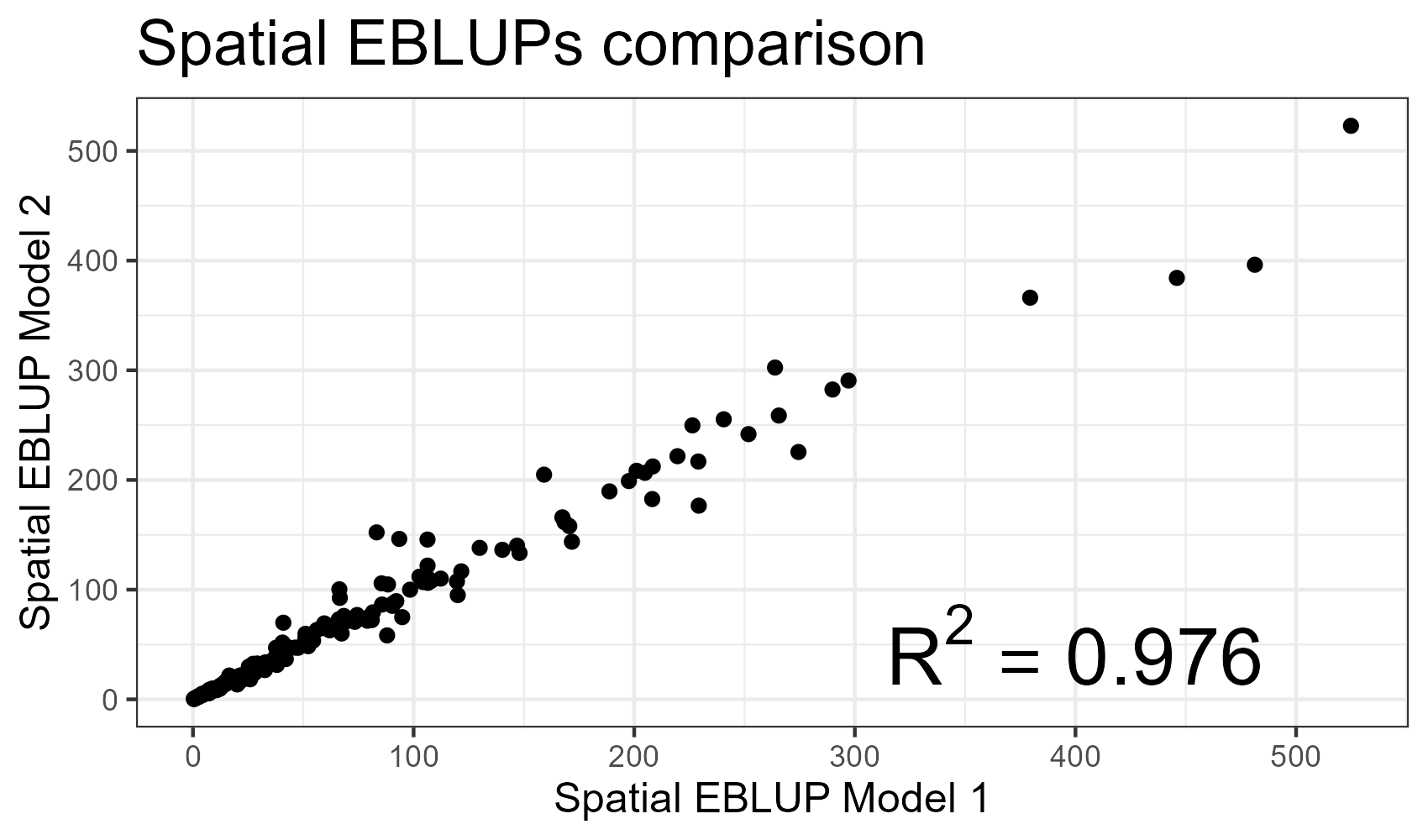}\\
\includegraphics[width=5.1cm]{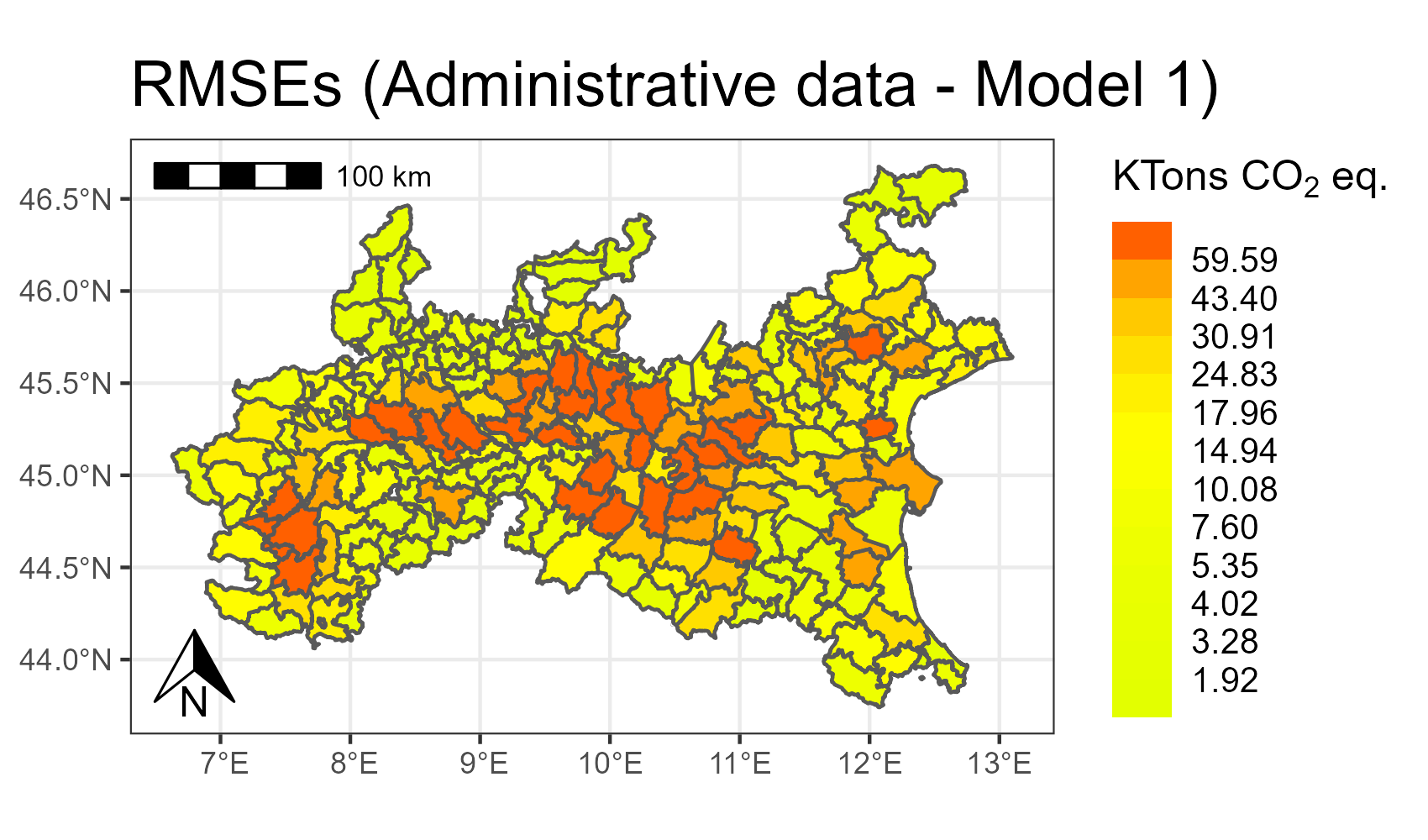}
\includegraphics[width=5.1cm]{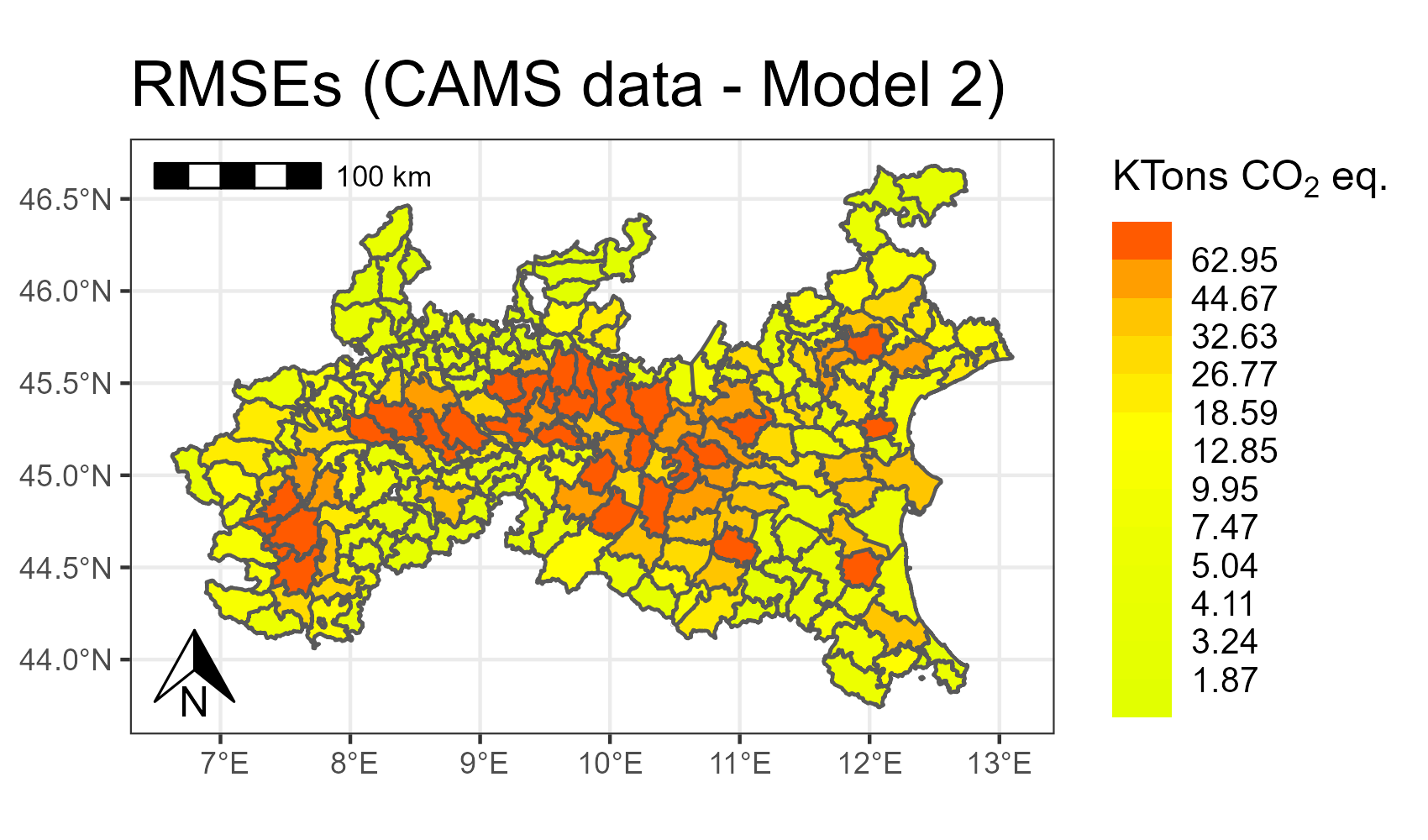}
\includegraphics[width=4.7cm]{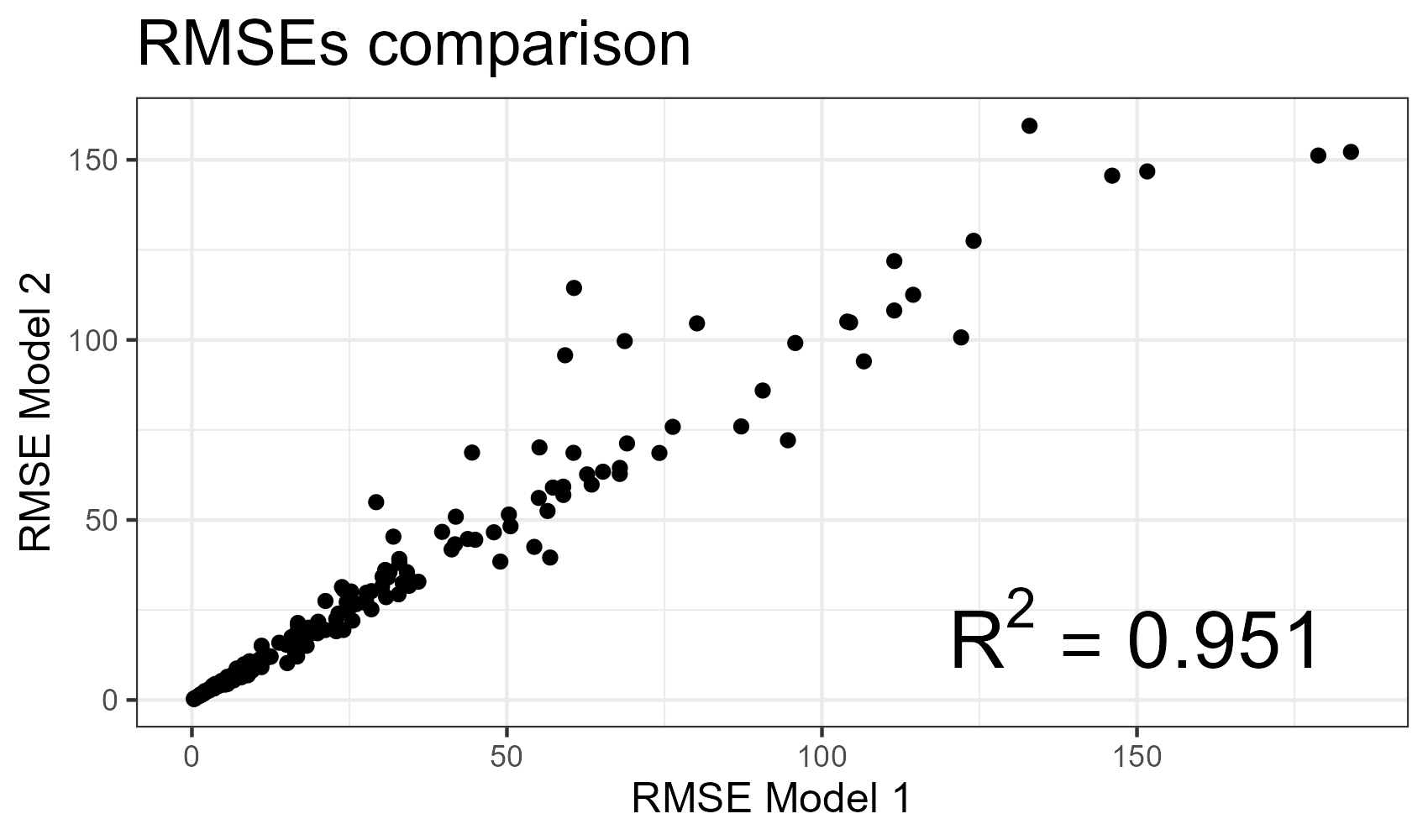}
\caption{Spatial Fay–Herriot model with SAR random effects. Left panels: (a-d) ASR-level EBLUPs of agricultural CF and corresponding RMSEs from Model 1 (KTons CO$_2$ eq.). Central panels: (b-e) ASR-level EBLUPs of agricultural CF and corresponding RMSEs from Model 2 (KTons CO$_2$ eq.). Right panels: (c-f) Scatterplots comparing EBLUPs and RMSEs obtained from Model 1 and Model 2 (KTons CO$_2$ eq.).}
\label{fig:EBLUPandRMSE}
\end{figure}

A graphical comparison between the two regression models is reported in Figure \ref{fig:EBLUPandRMSE}.

The two maps in the first row of the panel display the CF estimates obtained using Model 1 and Model 2. The two approaches, namely the spatial EBLUP based on census information and the spatial EBLUP based on satellite data, show a remarkable degree of similarity, as further illustrated by the scatterplot in the same panel. The squared correlation coefficient between the two sets of predictions is as high as 0.976, with moderately larger differences observed only in the highest values. Similarly, the RMSE of the two approaches appears to be comparable, both in terms of spatial structure and the magnitude of the prediction error, as clearly illustrated in the second row of graphs in the panel in Figure \ref{fig:EBLUPandRMSE}. It is also noteworthy that the largest values of the RMSE are observed to correspond to the largest values of the punctual estimates across all the models.

\begin{figure}
\centering
\includegraphics[width=16cm]{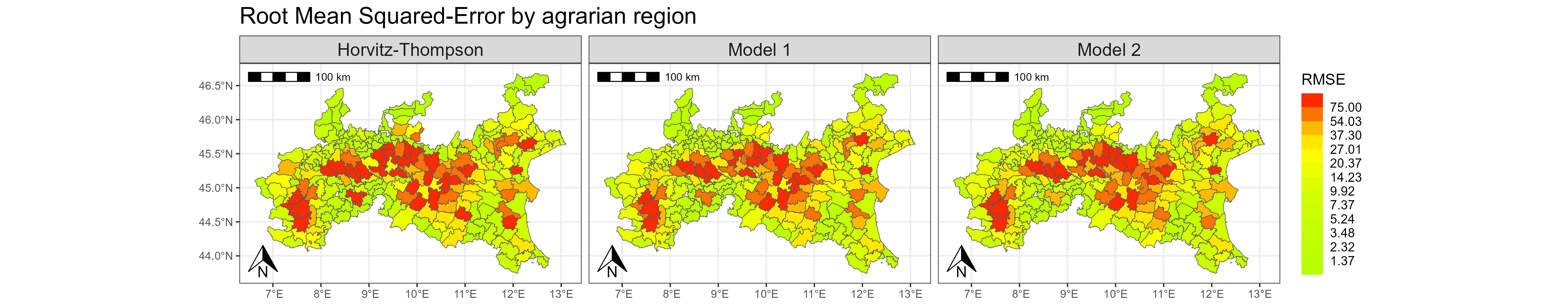}
\caption{Spatial Fay-Herriot model with SAR random effects. Left panel: (a) Estimated RMSE for the direct HT estimator. Central panel: (b) Estimated RMSE for the EBLUPs in Model 1. Right panel: (c) Estimated RMSE for the EBLUP in Model 2. Colors are on the same scale to provide are fair comparison.}
\label{fig:MAPrmse}
\end{figure}

It should be noticed that in Figure \ref{fig:EBLUPandRMSE} we represented the EBLUPs and the RMSEs in their original scale. However, to provide a fair comparison, in Figure \ref{fig:MAPrmse} we report the estimated RMSEs for the direct HT estimator and for the two models in the same scale. First, the maps clearly show the remarkable improvements we obtain moving from a direct estimation to a model-based estimation; second, relatively to the SFH models, they also show that the use of multiple census auxiliary information (\textit{Model 1}) and a single satellite auxiliary information (\textit{Model 2}), which actually synthesizes the most relevant spatio-temporal dynamics of the CF across the regions, return very close point estimates and variability estimates. However, as stressed earlier, while satellite data are available on a (almost) real-time basis, other kind of information might be not available and properly updated.

Recall that the variance of the HT estimators can be computed as in Equation \ref{eq:HTvar}, while the variance (and its square root) of the EBLUPs is computed via the analytical expression by \cite{singh2005spatio} in Model 1 and via the double bootstrap procedure (Algorithm \ref{AlgDoubleBoot}) in Model 2. 
Figure \ref{fig:MAPrmse} corroborates the benefits of a model-based approach, as evidenced by the lower variability of the EBLUPs compared to the direct estimator. Indeed, Fay-Herriot modeling leads to a reduction greater than $20\%$ in terms of the coefficient of variability on average, with the greatest improvement in areas where the population size and the sampling rate are minimal.

\subsection{Diagnostics check and robustness analysis}
While direct estimators rely solely on survey data and corresponding design weights, Fay–Herriot model-based estimates rest on the standard assumptions of linear mixed models. Specifically, both sampling errors and random effects (see Section~\ref{sec:SAE}) are assumed to follow a Gaussian distribution. In addition, we remark that the bootstrap methodology proposed in this paper also relies heavily on the Gaussian assumption. Verifying these assumptions is therefore essential to ensure the reliability of the results obtained from the statistical analysis.

The main outcomes of an in-depth diagnostic on the appropriateness of the conditional simulation analysis are reported in the Appendix \ref{App_C}, indicating that the approach here implemented is adequate and can be successfully applied. Hereinafter, we present the main results obtained from the diagnostic analysis of the parametric bootstrap outcomes.

\begin{figure}
\centering
\includegraphics[width=12cm]{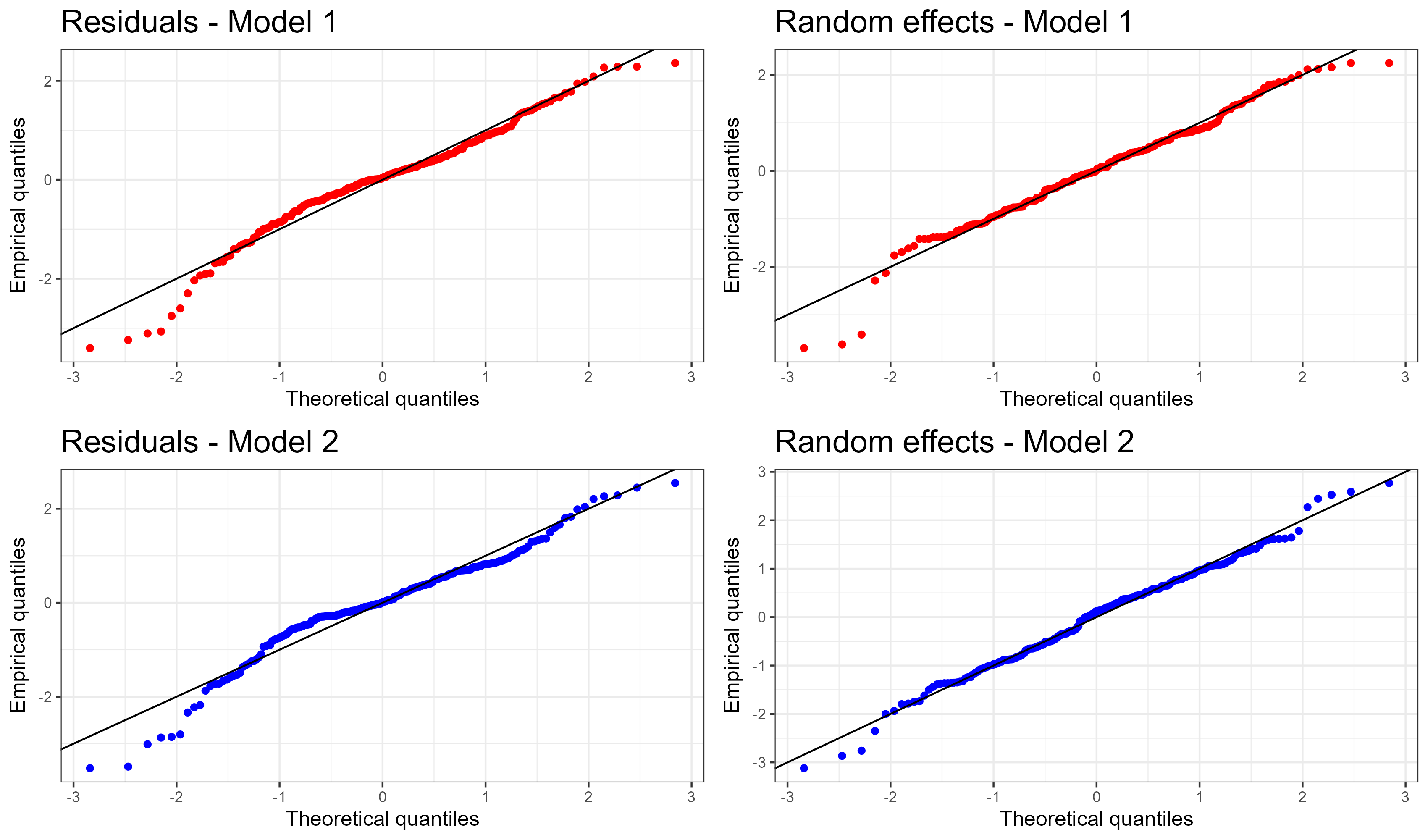}
\caption{Spatial Fay-Herriot model with SAR random effects. Left panels: (a-c) Q-Q normal plots for the model residuals. Right panels: (b-d) Q-Q normal plots for estimated SAR random effects.}
\label{fig:Diagnostics}
\end{figure}

In Figure~\ref{fig:Diagnostics}, the Q–Q normal plots for the sampling errors (left panels) and the random effects (right panels) obtained from Model~1 (first row) and Model~2 (second row) are presented. The empirical standardized quantiles are reported on the $y$-axis, while the theoretical quantiles from a Gaussian distribution are shown on the $x$-axis. Across all four plots, the normality assumption appears largely reasonable - particularly for the random effects - thus supporting the Gaussian specification underlying the models. Nonetheless, some deviations from normality persist. In particular, residuals from both Model~1 and Model~2 exhibit a mildly heavier left tail, as indicated by points falling below the reference line. These potential tail-related outliers correspond to ASRs located in central areas characterized by intensive, large-scale agricultural activities, where the survey includes only a limited number of farms despite a substantially larger farm population recorded in the census. For example, the ASR \textit{Pianura del Canale Cavour}, with $533$ census-level farms only $9$ of which are included in the sample, presents the highest departure from normality in the residuals on the lower tail in Model 2.

In addition, we compare the estimated CF values obtained from our models with those derived from the direct estimates at the appropriate spatial level. Recall that the FADN survey is designed to be statistically representative only at the NUTS-2 regional level and not at lower spatial disaggregation. However, each ASR belongs to one and only one NUTS-2 region within the study area, namely: Piedmont, Lombardy, Veneto, and Emilia-Romagna. Thus, we compared the CF estimates for these four NUTS-2 regions as provided by the HT estimator and by our model. In the latter case, CF estimates were first computed at the ASR level using the proposed model and subsequently aggregated to the NUTS-2 level. The results, reported in Table~\ref{tab:NUTS2est}, show that the model-based approach yields estimates that are very close to the direct ones, once again confirming the appropriateness of the proposed methodology. The model proves to be consistent with the direct estimates when the latter are properly applied, while providing reliable results at a finer spatial resolution.

\begin{table}
\centering
\small
\begin{tabular}{@{}lccc@{}}
  \toprule
 Region (NUTS-2) & Direct est. & Model 2 & $95\%$ bootstrap percentile CI \\  
  \midrule
 Emilia Romagna & 3,179 & 2,542 & (1,041; 7,997) \\ 
 Lombardy & 6,754 & 4,730 & (2,551; 24,079) \\ 
 Piedmont & 3,654 & 3,726 & (1,109; 9,601) \\ 
 Veneto & 1,818 & 2,466 & (1,965; 17,025) \\ 
   \bottomrule
\end{tabular}
\caption{Estimated CF at the NUTS-2 level using both direct and model-based approaches. Confidence intervals are obtained by aggregating model predictions at the NUTS-2 level within each bootstrap replicate and computing bootstrap percentile confidence intervals.}
\label{tab:NUTS2est}
\end{table}

Finally, to assess the impact of spatial correlation in the data, we also estimated the standard Fay--Herriot model without incorporating spatial dependence, that is, assuming that the area effects are independently drawn from a normal distribution. For brevity, the corresponding results are not reported in the main text but are available in Appendix \ref{App_A}. Overall, the signs and magnitudes of the regression coefficients obtained under the non-spatial model were broadly consistent with those of the spatial Fay--Herriot specification. As might be expected, once spatial correlation is accounted for, some differences emerge in terms of statistical significance, although both models identify the same subset of explanatory variables as significant.

\section{Discussion and final remarks} \label{sec:conclusions}
In this paper, we investigated the spatial dynamics of the carbon footprint (CF) associated with farms in the Po Valley, in Northern Italy, using small area models that exploit several spatially variable covariates. Data on CF and techno-economic specialization at the farm level were provided by the FADN program, the European sample survey system that collects detailed structural and accounting data on agriculture and livestock. Among others, a farm-specific greenhouse gas (GHG) emission indicator, namely the agricultural carbon footprint indicator, was used as proxy of the environmental impact of agricultural activities.

To get reliable estimates of the agricultural CF, we considered the Fay-Herriot model with spatially correlated random effects \cite{PratSal2008}. Specifically, we compared two non-nested empirical specifications. The first, previously referred to as Model 1, augments the regression component with area-level information on land use and land cover, agricultural labor, orography, and the number of breeding animals; the second specification, previously referred to as Model 2, relies on a single covariate derived from satellite-based measurements of ammonia emissions, which serve as an aggregate but consistent proxy for the environmental impact of agricultural and livestock farming activities. The first specification relies on census information collected by institutions through census (e.g., utilized agricultural area provided by the Italian National Institute of Statistics) or administrative surveys (e.g., the veterinary database for animals); such data are often collected across long temporal spans without a cyclical update. The satellite specification, instead, relies on continuously-monitored measurements which cover large cross-national areas, such as the European continent, and are updated several times per year.

The issue of spatial misalignment between gridded satellite data and the administrative boundaries of agrarian regions is addressed through a geostatistics-based algorithm integrated with a bootstrap procedure. This methodological framework enables to account for the propagation of uncertainty arising from the data alignment process toward the estimation of model parameters, thereby ensuring the reliability of the inferential results.

Empirical results clearly showed that the small area model-based approach significantly outperforms direct estimators that do not use auxiliary information. The CF was found to be very heterogeneous among the agricultural sub-regions, while remaining coherent with the economic and environmental configuration of Northern Italy. In particular, the CF is very high in the plain areas, where intensive livestock farming is widespread, confirming and extending the other empirical studies on the topic \cite{CarilloEtAl2024, MarongiuEtAl2022, MarongiuEtAl2024, OttoEtAl2024}.

Furthermore, the findings reveal a strong correlation and spatial concordance between the census-based and satellite-based models, which produce mutually consistent estimates of the agricultural carbon footprint. In line with results reported in other research domains (see, among others, \cite{PermatasariEtAl2025, edochie2025small, NewhouseEtAl}), this evidence suggests that, at least in the context of spatial prediction based on survey data, satellite-derived measurements can serve as a reliable alternative to census and administrative data. This, in turn, enables meaningful cross-country comparisons where relevant, while simultaneously reducing data retrieval, preprocessing, and computational costs.

\section*{Acknowledgments}
We deeply acknowledge the SCARFACE (Sequestering CARbon through Forests, AgriCulture, and land usE) project, funded by the University of Milano-Bicocca, and the AgroGeoStat project, funded by the University of Milano-Bicocca and the \textit{Centro di ricerca Politiche e Bioeconomia, Consiglio per la ricerca in agricoltura e l'analisi dell'economia agraria (CREA-PB)} , for their precious contribution to the study by providing and illustrating the data.\\
Finally, we greatly acknowledge the DEMS Data Science Lab for supporting this work by providing computational resources.


\bibliography{references}

\bibliographystyle{apalike}

\newpage
\section*{Appendix A: Comparison between Fay-Herriot model with independent random effects and with spatially-correlated random effects}
\label{App_A}
To assess the efficacy of the small area estimators, we undertake a comparison of the root mean squared error (RMSE) at the area level obtained by the Fay-Herriot model with independent random effects (FH) and with spatially-correlated random effects (SFH).

Table \ref{tab:FH_merged_coefficients} reports the estimated regression coefficients of the SFH (columns 2 and 3) and FH (columns 4 and 5) models when considering the set of census information (columns 2 and 4) and the single satellite variable (columns 3 and 5). It is straightforward to notice that magnitudes, signs and significances are preserved in almost any cases, with some negligible difference (e.g., standard working days and bovines decrease their level of significancy when the spatial correlation is embedded).

\begin{table}[ht]
\centering
\small
\caption{Estimated regression coefficient of the Fay–Herriot models with independent random effects (columns 2 and 3) and with spatially-correlated random effects (columns 4 and 5). Standard errors are provided under parentheses. The covariates included in the linear predictor of both models have been standardized.}
\label{tab:FH_merged_coefficients}

\begin{tabular}{lcccc}
\toprule
& \multicolumn{2}{c}{\textbf{Spatial Fay-Herriot}} & \multicolumn{2}{c}{\textbf{Fay-Herriot}} \\
\midrule
\textbf{Variable} 
& \textbf{Model 1} & \textbf{Model 2} 
& \textbf{Model 1} & \textbf{Model 2} \\
\midrule
\multirow{2}{*}{Intercept} 
  & $3.454$ $^{***}$ & $3.451$ $^{***}$ 
  & $3.482$ $^{***}$ & $3.480$ $^{***}$ \\
  & \footnotesize{($0.086$)} & \footnotesize{($0.227$)} 
  & \footnotesize{($0.086$)} & \footnotesize{($0.091$)} \\
\multirow{2}{*}{Utilized agricultural area} 
  & $0.587$ $^{***}$  & -- 
  & $0.587$ $^{***}$ & -- \\
  & \footnotesize{($0.138$)} & 
  & \footnotesize{($0.119$)} & \\
\multirow{2}{*}{Standard working days} 
  & $-0.342$ $^{**}$ & -- 
  & $-0.561$ $^{***}$ & -- \\
  & \footnotesize{($0.117$)} & 
  & \footnotesize{($0.120$)} & \\
\multirow{2}{*}{Total bovine heads} 
  & $0.467$ $^{**}$ & -- 
  & $0.601$ $^{***}$ & -- \\
  & \footnotesize{($0.182$)} & 
  & \footnotesize{($0.202$)} & \\
\multirow{2}{*}{Total swine heads} 
  & $-0.108$ & -- 
  & $-0.016$ & -- \\
  & \footnotesize{($0.180$)} & 
  & \footnotesize{($0.196$)} & \\
\multirow{2}{*}{Average altitude} 
  & $0.161$ & -- 
  & $0.045$ & -- \\
  & \footnotesize{($0.181$)} & 
  & \footnotesize{($0.102$)} & \\
\multirow{2}{*}{NH$_3$ emissions} 
  & -- & $0.609$ $^{***}$ 
  & -- & $0.557$ $^{***}$ \\
  & & \footnotesize{($0.120$)} 
  & & \footnotesize{($0.091$)} \\
Spatial autocorrelation $\rho$ & $0.717$ & $0.734$ & / & / \\
\bottomrule
\end{tabular}

\end{table}

\begin{figure}
\centering
\includegraphics[width=5.1cm]{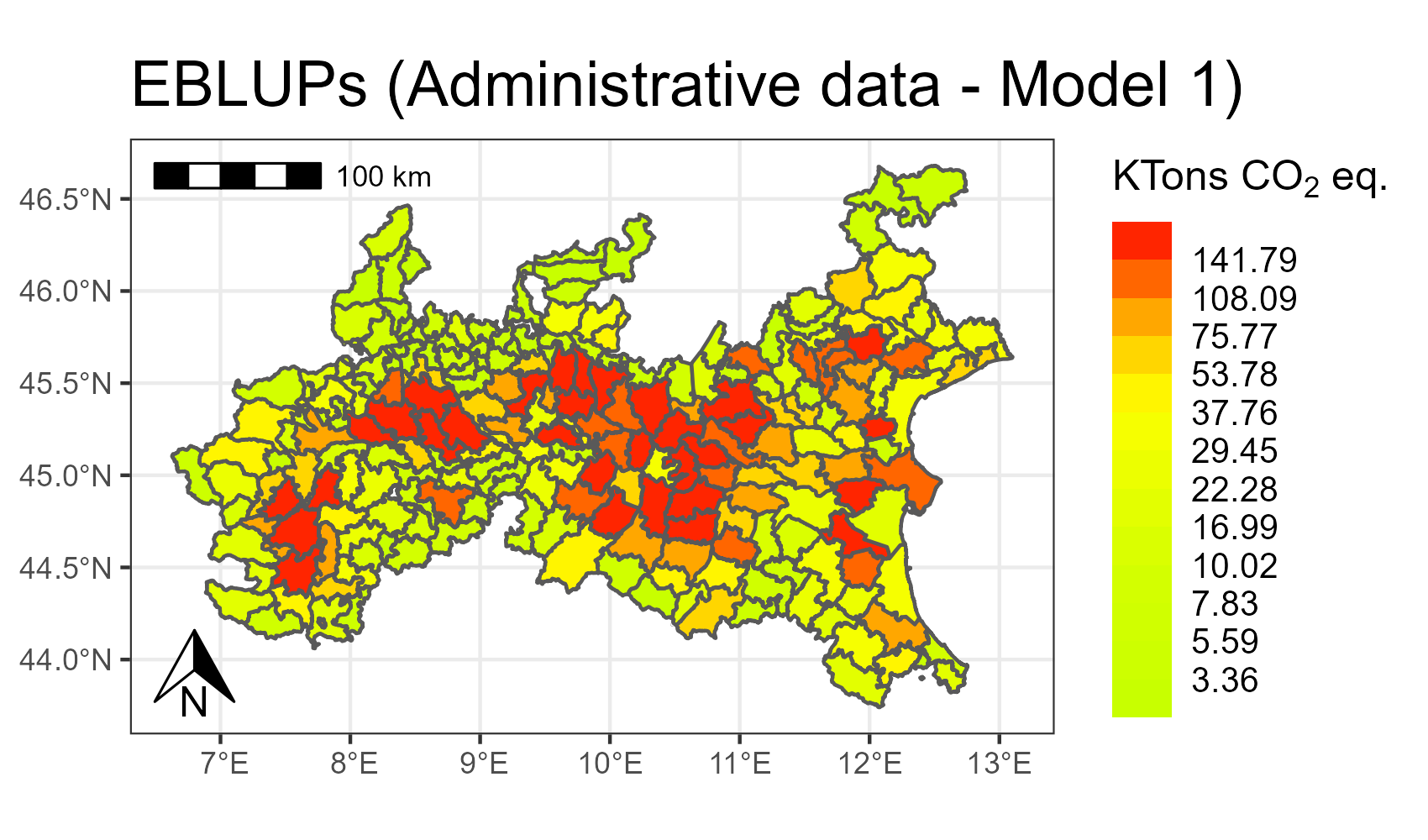}
\includegraphics[width=5.1cm]{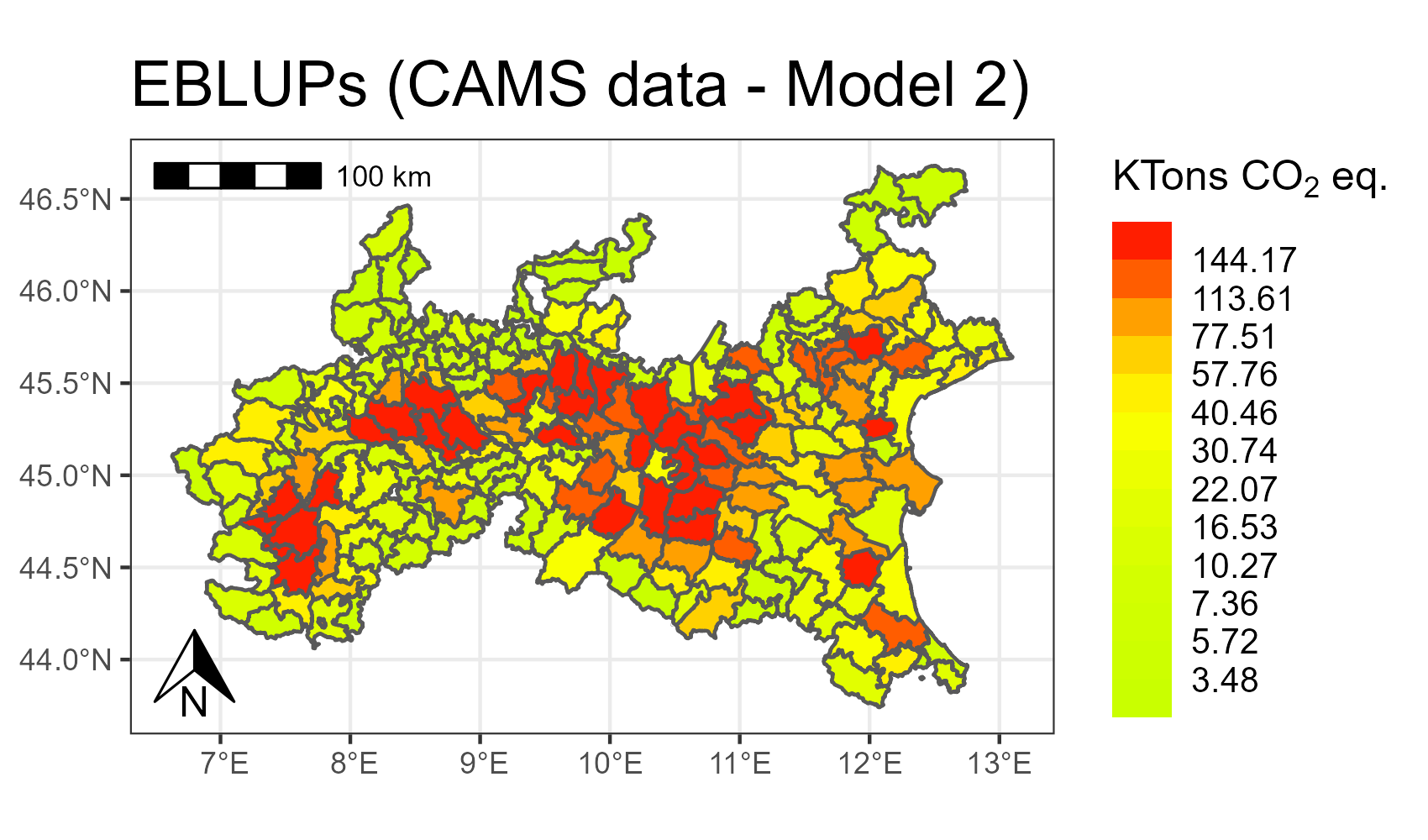}
\includegraphics[width=4.7cm]{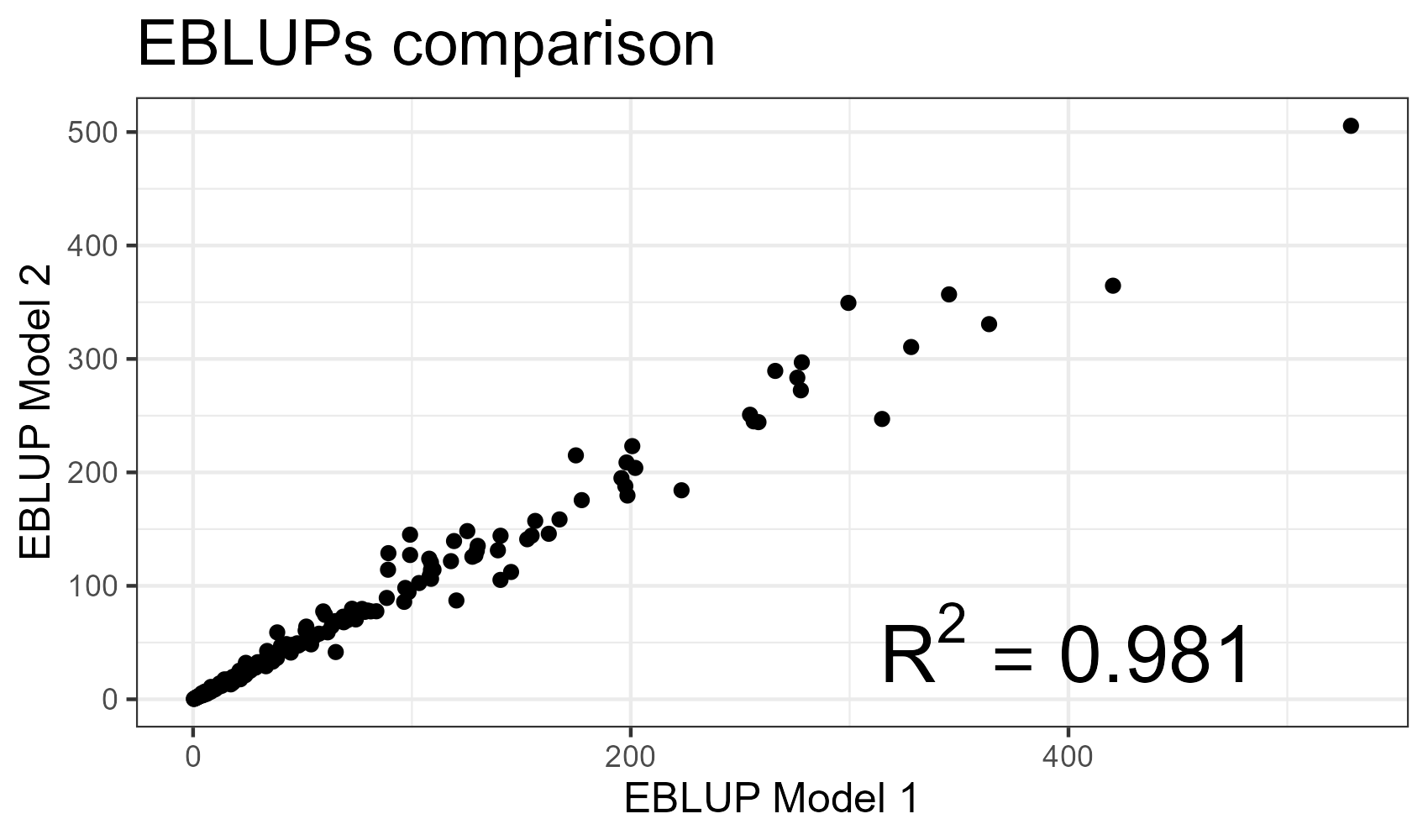}\\
\includegraphics[width=5.1cm]{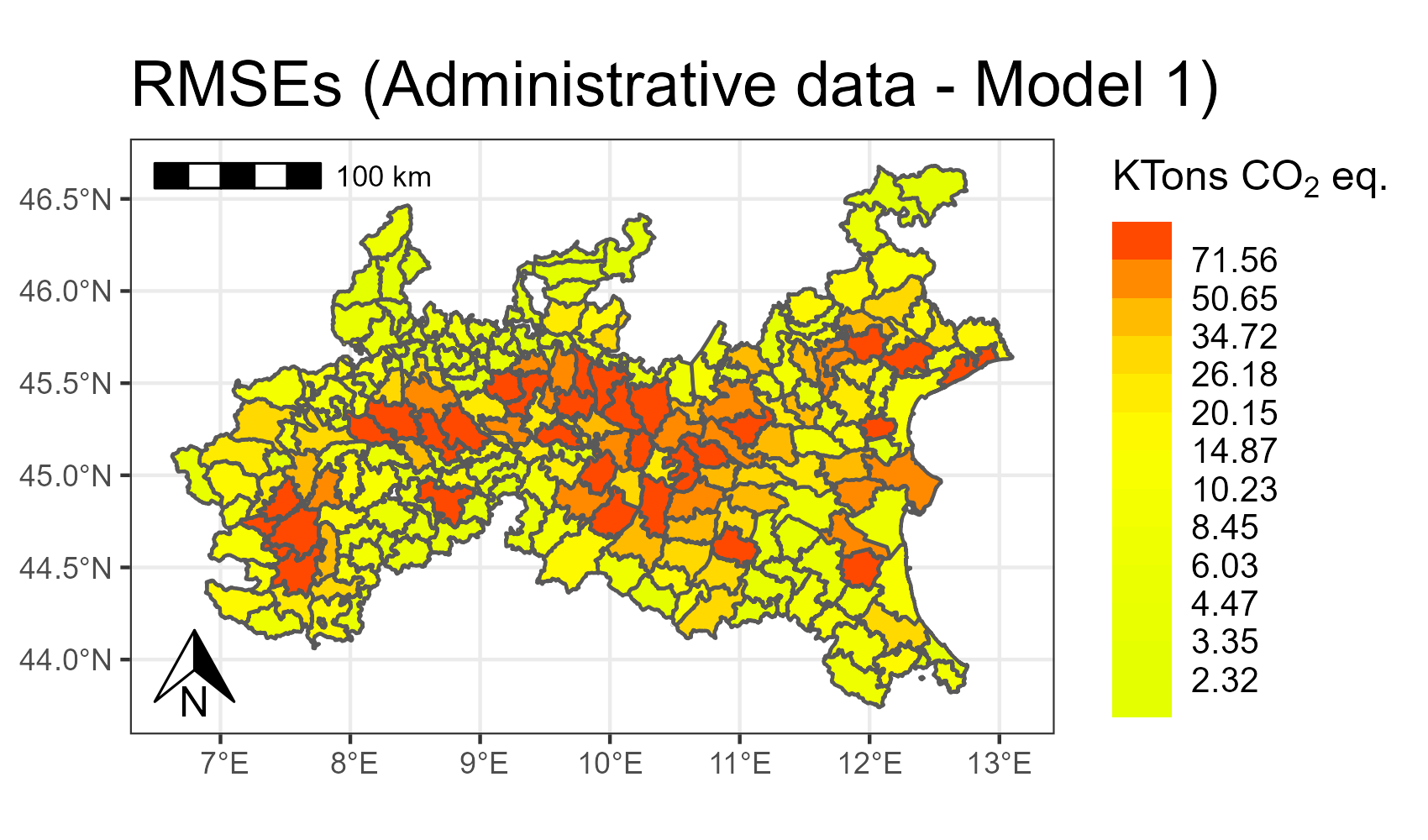}
\includegraphics[width=5.1cm]{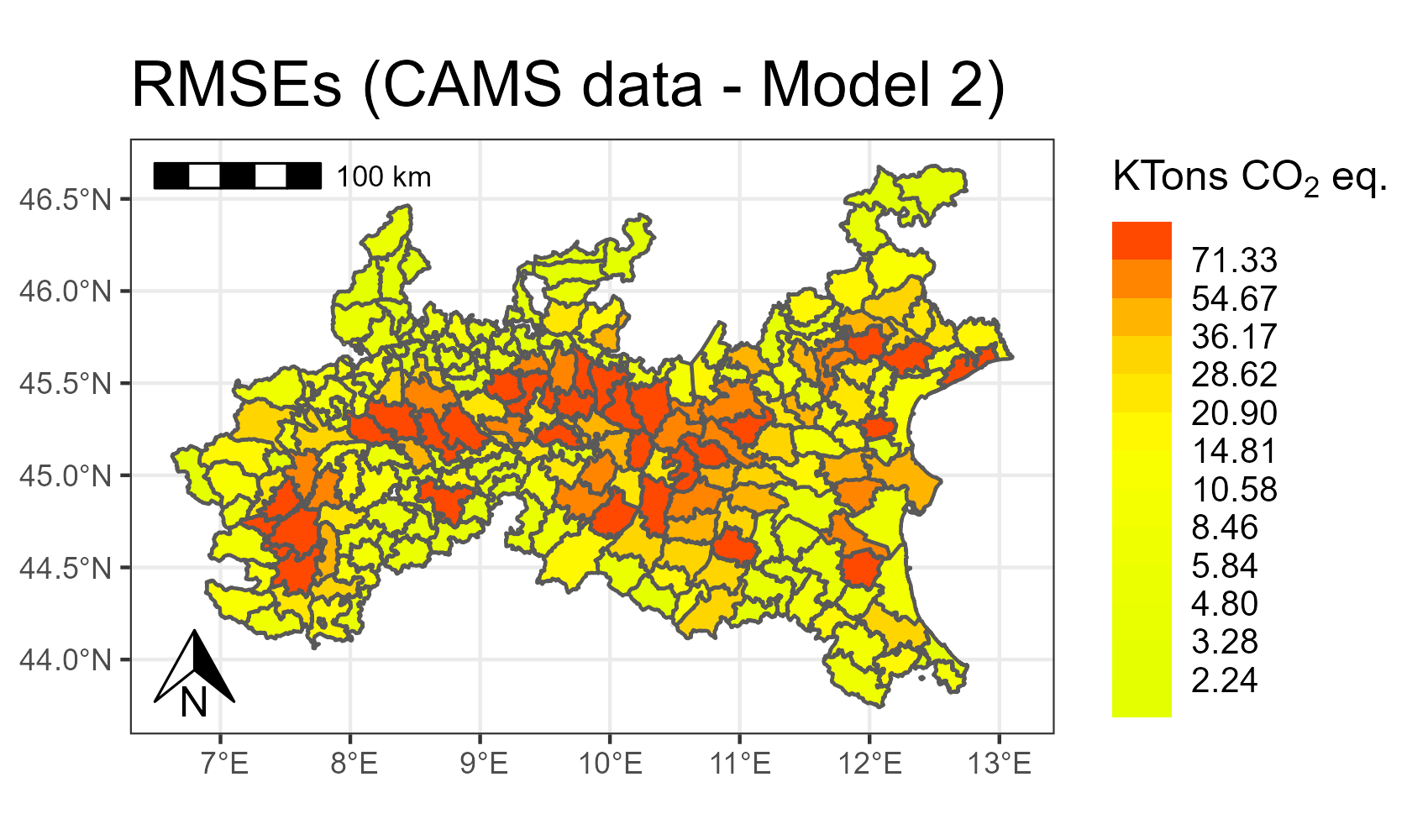}
\includegraphics[width=4.7cm]{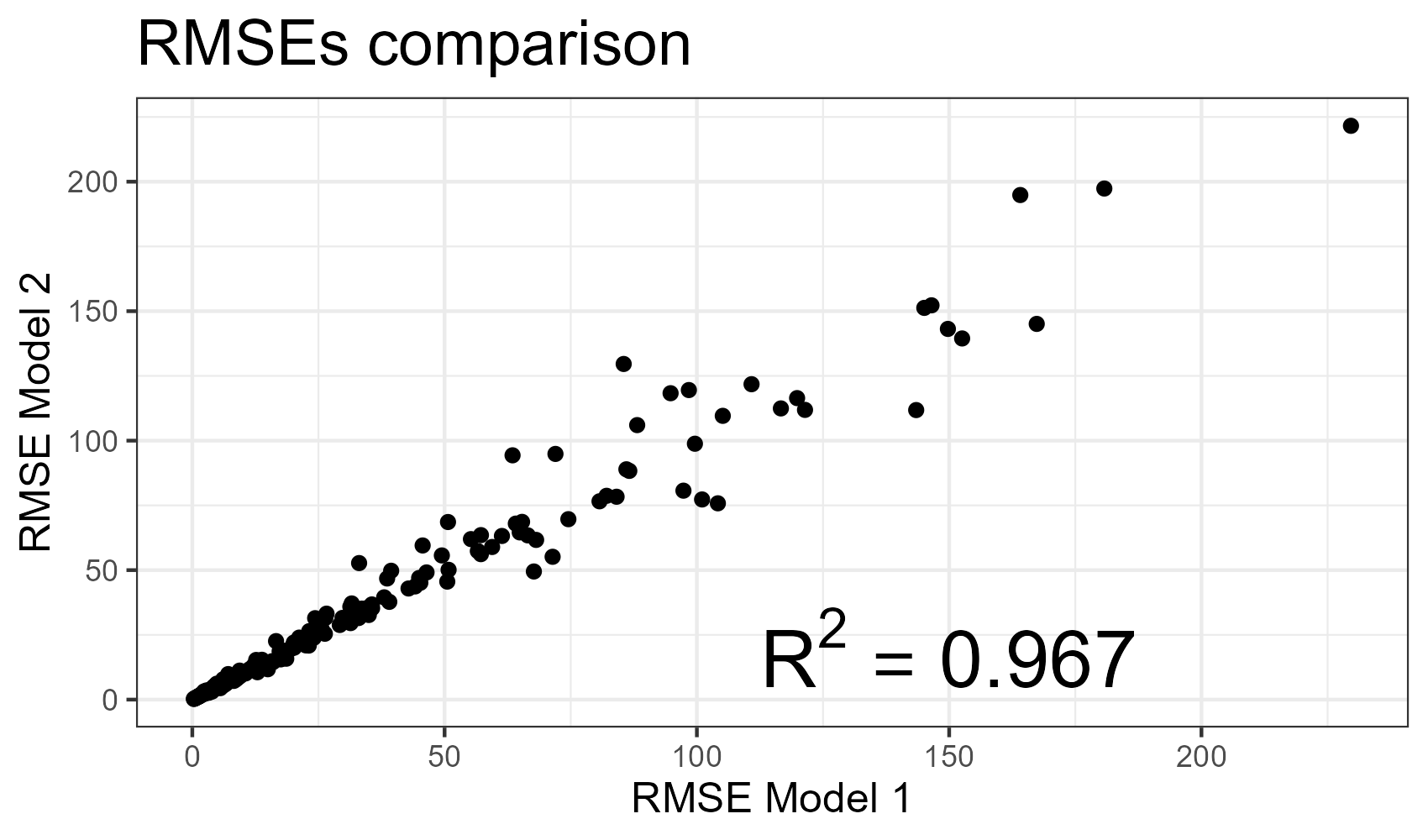}
\caption{Fay–Herriot model with independent random effects. Left panels: (a-d) ASR-level EBLUPs of agricultural CF and corresponding RMSEs from Model 1 (KTons CO$_2$ eq.). Central panels:(b-e) ASR-level EBLUPs of agricultural CF and corresponding RMSEs from Model 2 (KTons CO$_2$ eq.). Right panels: (c-f) Scatterplots comparing EBLUPs and RMSEs obtained from Model 1 and Model 2 (KTons CO$_2$ eq.).}
\label{fig:EBLUPandRMSEind}
\end{figure}

\begin{figure}
\centering
\includegraphics[width=16cm]{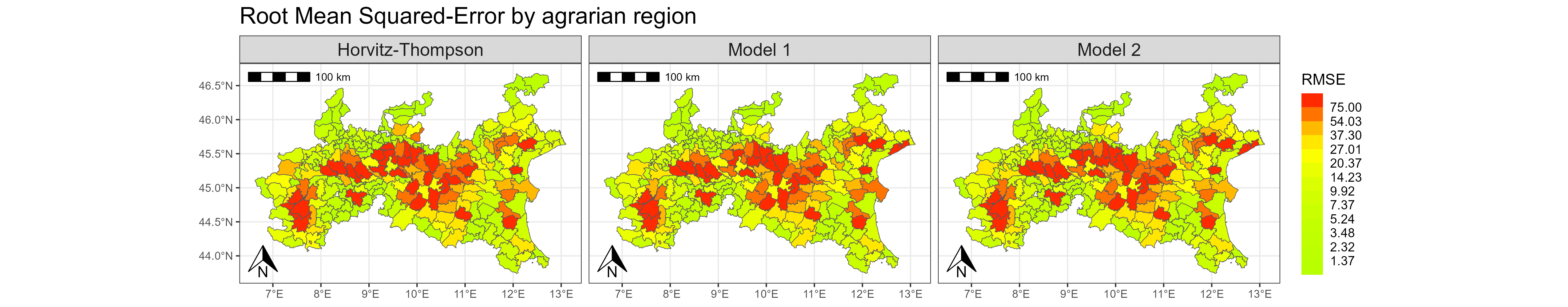}
\caption{Fay-Herriot model with independent random effects. Comparison of the variability of different approaches without spatial autocorrelation}
\label{fig:MAPrmseIND}
\end{figure}

\begin{figure}
\centering
\includegraphics[width=12cm]{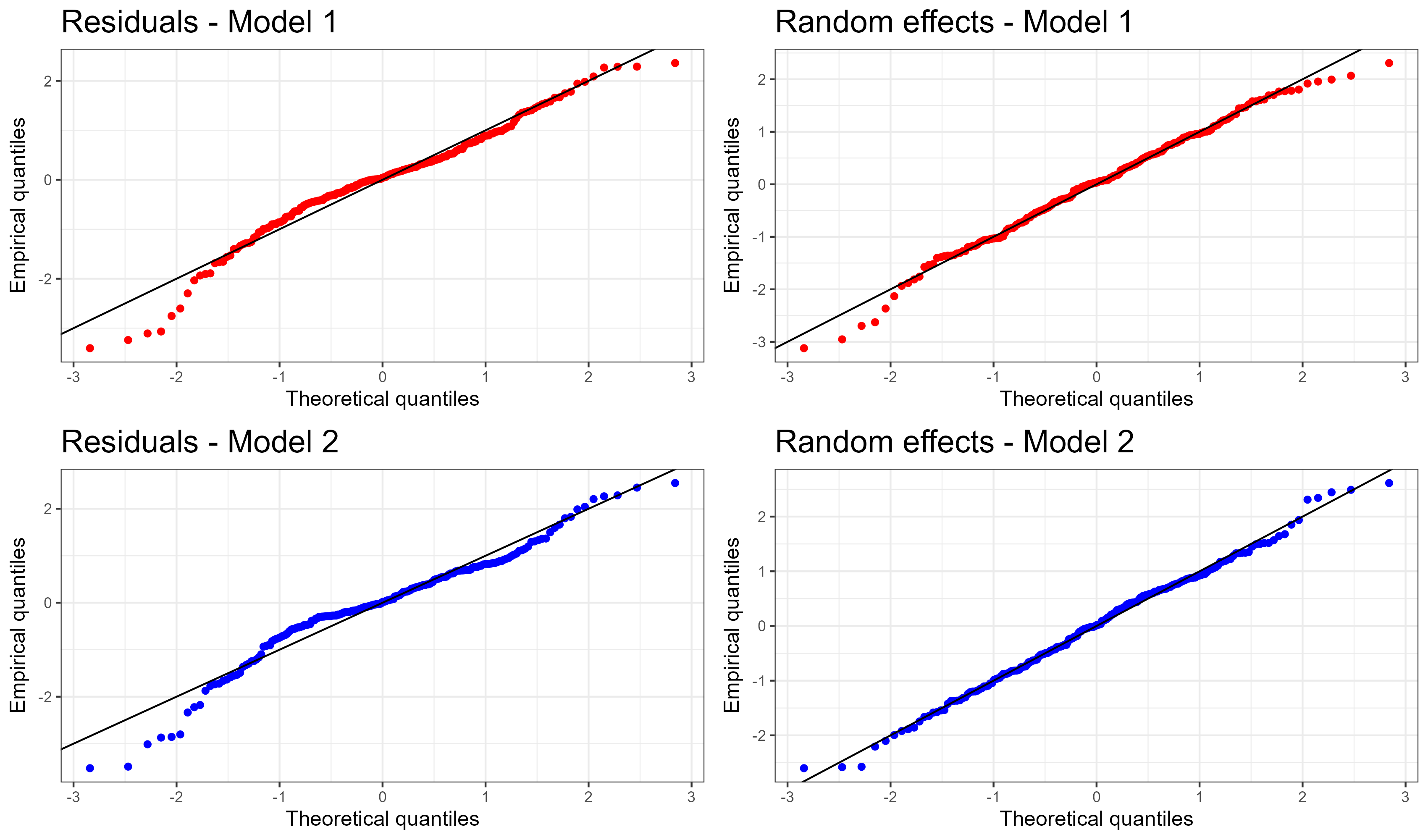}
\caption{Fay-Herriot model with independent random effects. Left panels: (a-c) Q-Q normal plots for the model residuals. Right panels: (b-d) Q-Q normal plots for estimated SAR random effects}
\label{fig:DiagnosticsIND}
\end{figure}

Further results for the FH model are available in Figures \ref{fig:EBLUPandRMSEind}, \ref{fig:MAPrmseIND}, and \ref{fig:DiagnosticsIND}. Summing up, the results suggest a very moderate difference between the FH and SFH models. Nevertheless, improvements in terms of variability of the estimates can be noticed.

\section*{Appendix B: Diagnostics check of geostatistics simulations } 
\label{App_B}
A key step of the bootstrap procedure described in Section~\ref{sec:SAE} relies on geostatistical simulations.

The simulations are performed using a Gaussian conditional approach, generating data from a Gaussian process with a variogram function defined by the Matérn model estimated from the actual dataset. 

Let $Y(s)$ denote the spatial process, $Y(s_1), \dots, Y(s_n)$ the observed data, and $Y^*(u)$ an unconditionally simulated value of the random field at location $u$.
The conditional simulation approach consists in simulating the value $Y_c(u)$ as
$$
  Y_c(u) = \hat{Y}(u) + \left[ Y^*(u) - \hat{Y}^*(u) \right],
$$
where $\hat{Y}(u)$ is the kriging prediction based on the real data, and
$\hat{Y}^*(u)$ is the kriging prediction based on the simulated values at the data locations.

$Y_c(u)$ is thus a \emph{conditional simulation} of the random field at location $u$, i.e., a random field consistent with both the observed data and the specified spatial model. In particular, if $u$ coincides with a sampling location $s_i$, then the simulated value equals the observed measurement.

Since conditional simulations replicate the observed data at each sampling location, we instead generate a number of measurement points equal to the CAMS locations (1259), sampled from a uniform distribution over the area of interest (i.e., the Po Valley). This ensures that, on average, each ASR is associated with the same number of measurement points as in the CAMS
dataset, while at the same time avoiding a mere replication of the observed data.

The proposed methodology relies heavily on the Gaussian assumption. Below we report the results of the detailed diagnostic that has been conducted to evaluate the validity of this assumption and, more broadly, the appropriateness of the overall procedure.

\begin{figure}
\centering
    \includegraphics[width=0.45\linewidth]{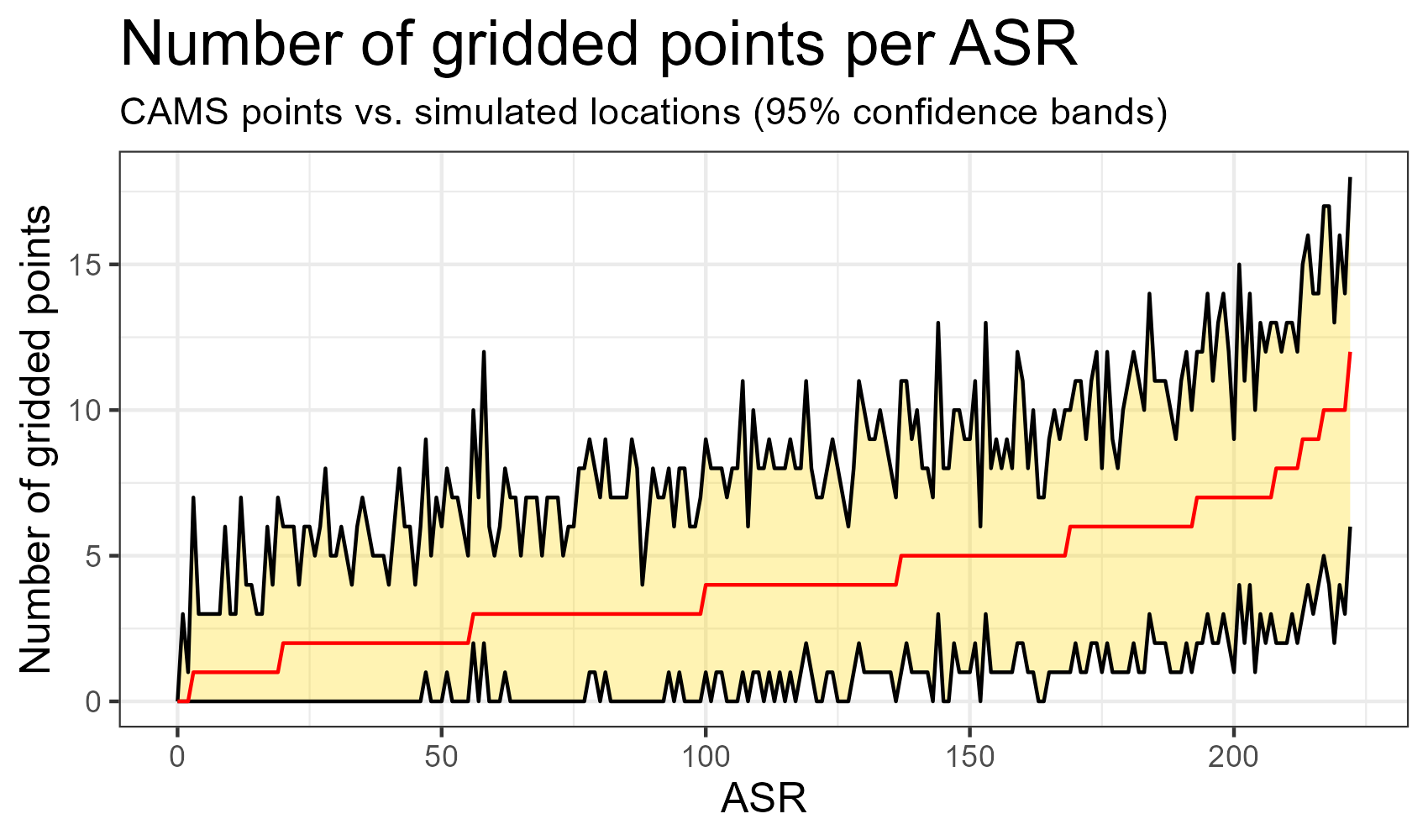}
    \includegraphics[width=0.45\linewidth]{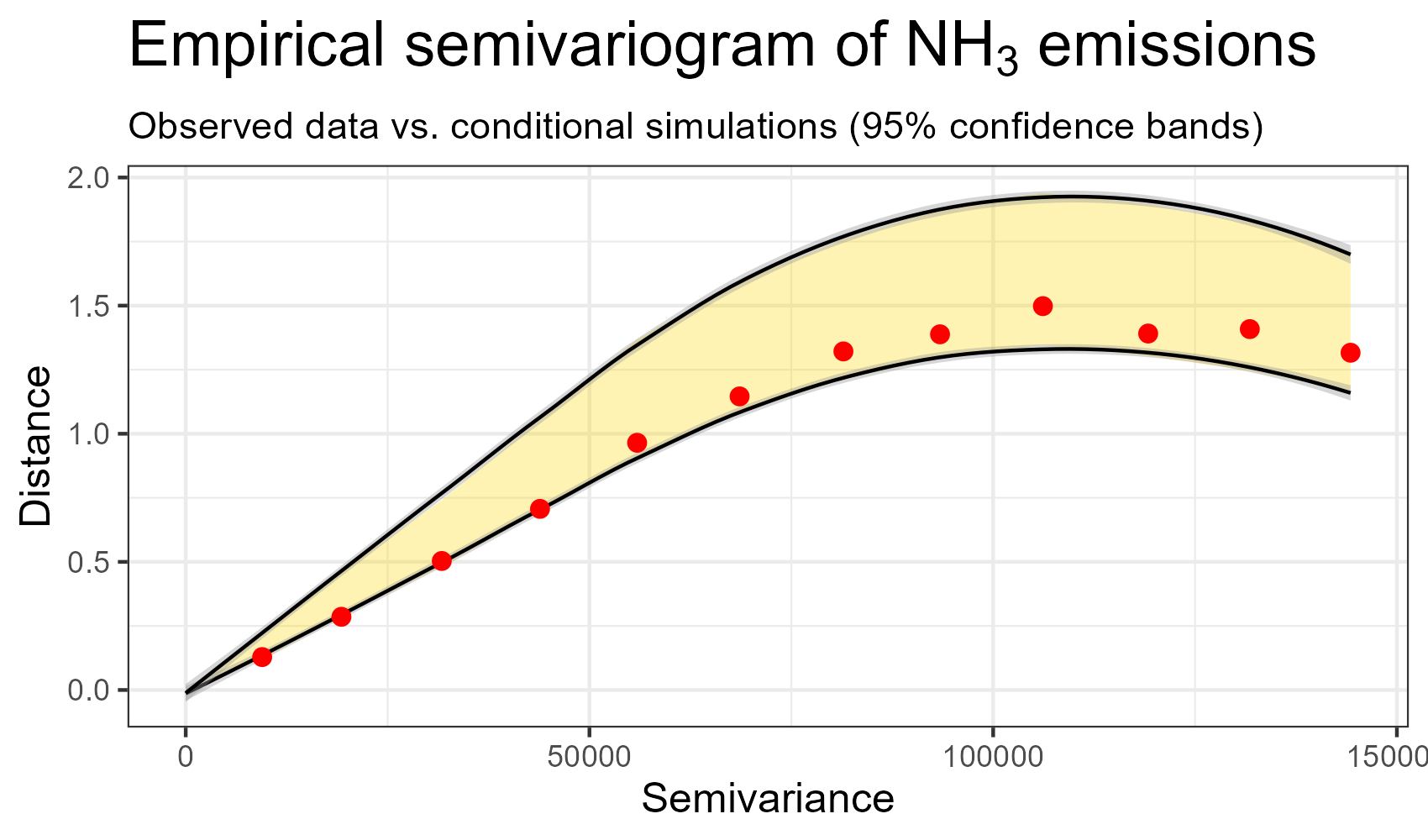}
\caption{Left panel: (a) Average number of simulated grid points per ASR. Right panel: (b) variogram envelope.}
\label{fig:Point_diagnostic}
\end{figure}

Figure \ref{fig:Point_diagnostic}-a shows the average number of simulated locations ($Y$-axis) plotted against the actual number of grid points present in each ASR in the CAMS dataset ($X$-axis). This plot demonstrates that the actual number of points per ASR is accurately reproduced by the simulation process.

Figure \ref{fig:Point_diagnostic}-b shows the envelope of the robust empirical variograms estimated for each simulated trajectory of the $log$-emission process (yellow area), together with the robust empirical variogram obtained from the actual data. 

This plot demonstrates that the spatial structure of the CAMS emissions is accurately reproduced by the simulation process.

The Monte Carlo envelopes were constructed using the 2.5th and 97.5th percentiles of the 1,000 simulated sample grids and trajectories.

\begin{figure}
\centering
 \includegraphics[width=0.45\linewidth]{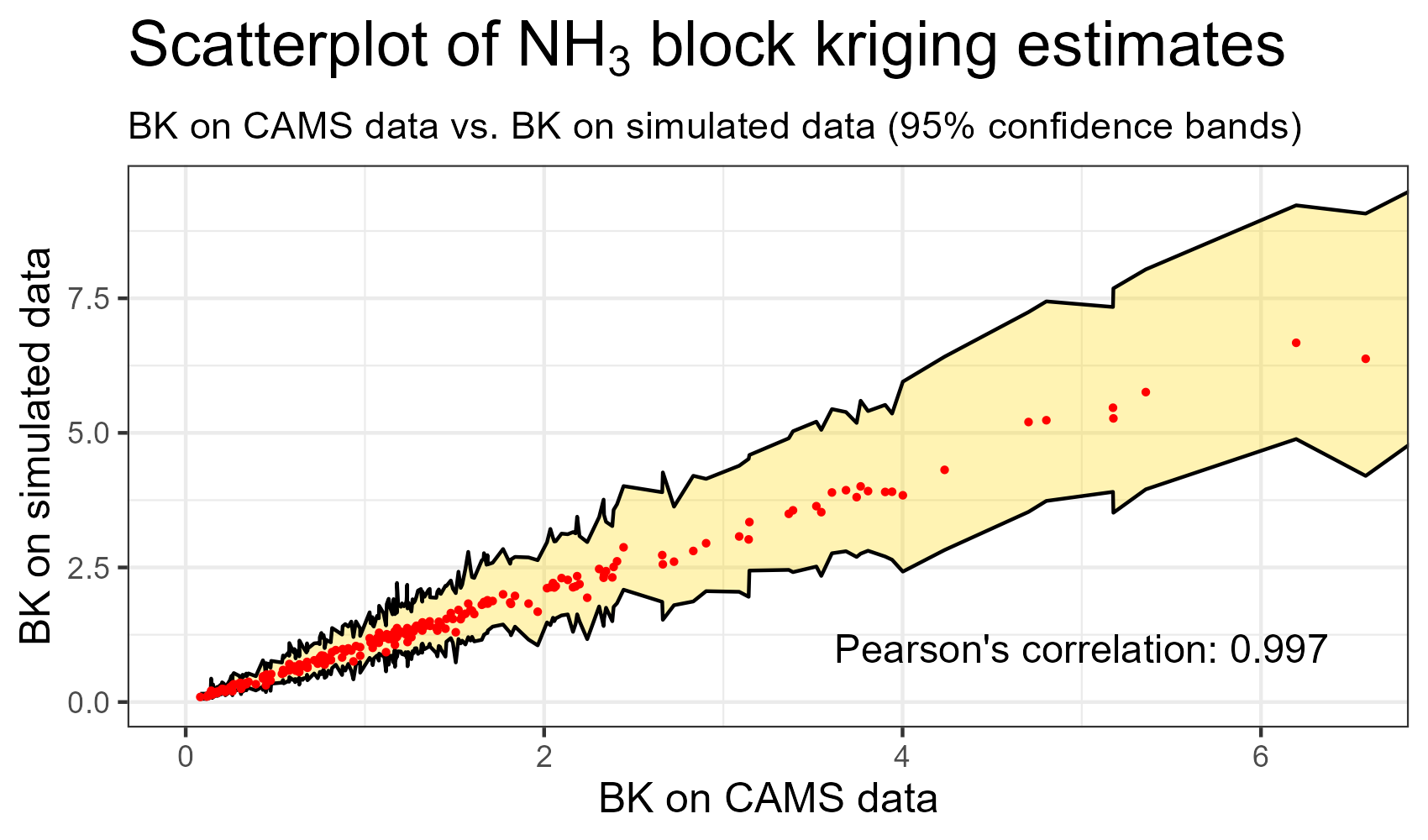}
 \includegraphics[width=0.45\linewidth]{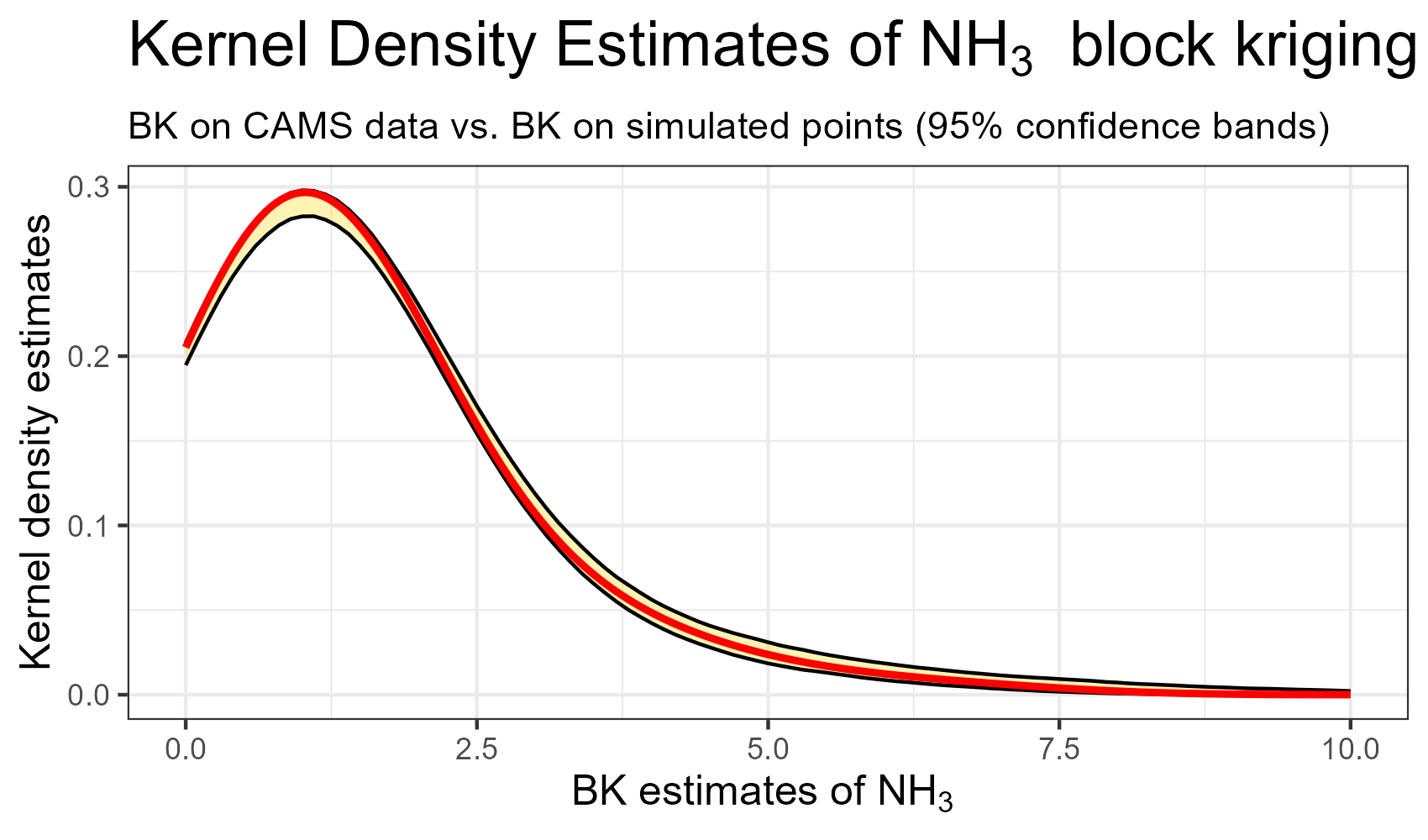}		
\caption{Left panel: (a) Average simulated ASR-level emission predictions plotted against the actual values.
Right panel: (b) Envelope of the spatial distributions of average simulated ASR-level predictions.}
\label{fig:CS_bk_diagnostic}
\end{figure}

Figure \ref{fig:CS_bk_diagnostic}-a shows the average block kriging predictions of the mean ASR emissions ($Y$-axis) plotted against the block kriging predictions obtained using the actual CAMS data ($X$-axis). Figure \ref{fig:CS_bk_diagnostic}-b shows the envelope of the kernel-smoothed spatial distributions of the block kriging predictions obtained for the 222 ASRs across all simulated trajectories (yellow area), together with the estimated distribution derived from the actual data. These plots demonstrate that the spatial structure of the block kriging predictions is accurately reproduced by the simulation process.

\section*{Appendix C: Diagnostics check of the bootstrap procedure}
\label{App_C}
The bootstrap procedure described in Section~\ref{sec:SAE} is parametric in nature and therefore relies on the underlying distributional assumptions.

Hereinafter, we present a set of diagnostics to assess the performance and robustness of the algorithm.

\begin{figure}
\centering
\includegraphics[width=0.45\linewidth]{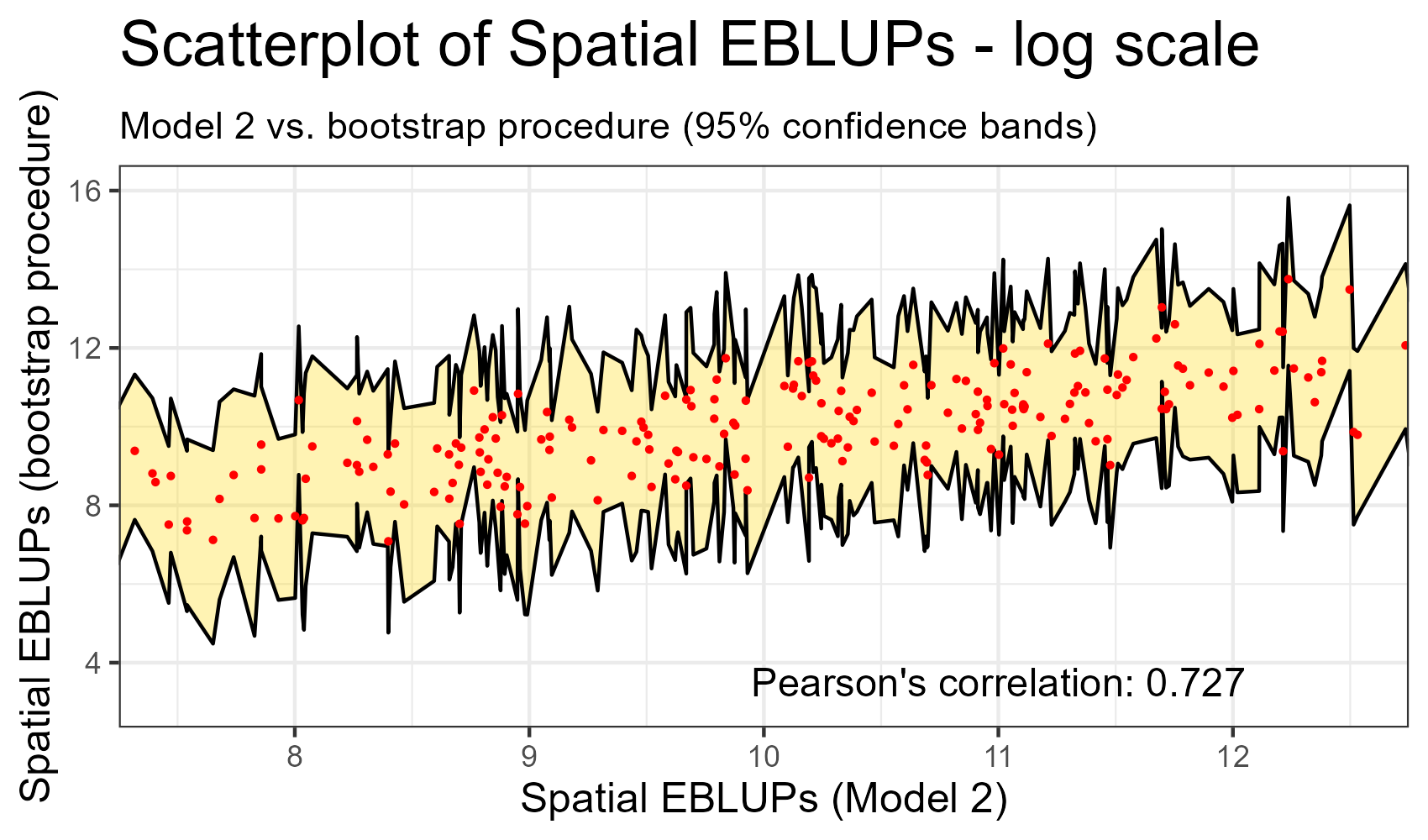}		
 \includegraphics[width=0.45\linewidth]{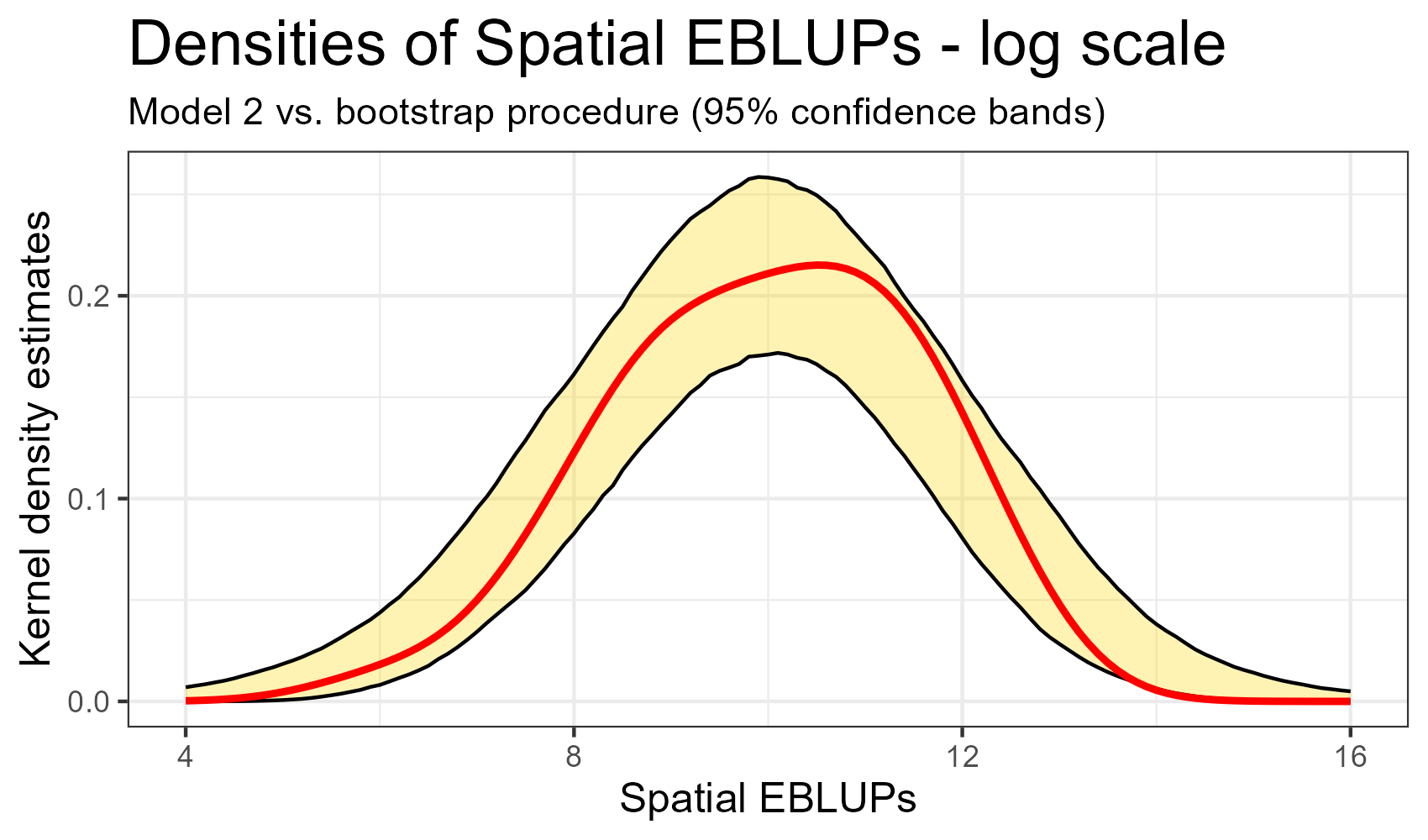}
\caption{Left panel: (a) Spatial predictions of total CF across bootstrap replicates on the log scale; red points represent the ASR-level bootstrap averages plotted against the EBLUPs obtained from the actual data.
Right panel: (b) Envelope of the spatial distributions of simulated CF totals per ASR on the log scale.}
\label{fig:boot_EBLUP_diagnostic}
\end{figure}

Figure~\ref{fig:boot_EBLUP_diagnostic}-a displays, in red, the average of the bootstrap prediction replicates of the carbon footprint total for each area, together with their $95\%$ envelope (on the $Y$-axis), plotted against the predictions obtained from the actual data (on the $X$-axis). The Monte Carlo envelope was constructed using the 2.5th and 97.5th percentiles of the 1,000 bootstrap replicates. The plot shows that, on average, the procedure is in strong agreement with the data, while properly accounting for the sampling variability represented by the bands.

Figure~\ref{fig:boot_EBLUP_diagnostic}-b shows the $95\%$ envelope of the kernel-smoothed spatial distributions of the predictions obtained from the 1,000 bootstrap replicates for the 222 ASRs considered (yellow area). The spatial distribution of the area-level predictions derived from the actual data is superimposed in red. This plot demonstrates that the bootstrap procedure preserves the spatial structure of the small area estimates of the total carbon footprint while appropriately capturing the sampling variability of these predictions.

Finally, Figure~\ref{fig:boot_MSE_diagnostic} displays the bootstrap prediction replicates of the CF total (on the $Y$-axis) for each ASR, together with their $95\%$ envelope, plotted against the MSE of the predictions obtained from the actual data (on the $X$-axis). The Monte Carlo envelope was constructed using the 2.5th and 97.5th percentiles of the 1,000 bootstrap replicates. All values are reported on the log scale for improved readability.
The plot shows that, on the one hand, the proposed procedure closely reproduces the prediction error of the model based on the actual data, thus not altering the prediction outcome. On the other hand, it highlights the clear advantages of using the SAE approach — even when relying on a single satellite-derived covariate — over the direct estimates of the carbon footprint.

\begin{figure}
\centering
\includegraphics[width=0.75\linewidth]{fig_BOOT_MSE.png}		
\caption{MSE differences between the spatial predictions of total CF per ASR obtained from the model using the satellite covariate and: the HT direct estimates (in green), the model-based spatial predictions obtained using administrative covariates (in blue), and the spatial predictions from the bootstrap replicates (yellow envelope). All values are reported on the log scale.}
\label{fig:boot_MSE_diagnostic}
\end{figure}

\end{document}